\documentclass[prd,aps,twocolumn,a4paper,floatfix,showpacs,footinbib]{revtex4-1}

\usepackage{times}
\usepackage{graphicx,psfrag}
\usepackage{mathrsfs}
\usepackage{amsmath,amsfonts,amssymb,amsthm}
\usepackage{pifont}
\usepackage{url}
\usepackage{comment}
\usepackage{xcolor}
\usepackage{float}
\usepackage{ulem}
\usepackage{MnSymbol}
\usepackage{enumerate}
\usepackage{upgreek}
\definecolor{dartmouthgreen}{rgb}{0.05, 0.5, 0.06}
\definecolor{deepcarmine}{rgb}{0.66, 0.13, 0.24}
\definecolor{deepsaffron}{rgb}{1.0, 0.6, 0.2}

\newcommand{\cmark}{\color{black}{\ding{51}}}%
\newcommand{\xmark}{\color{black}{\ding{55}}}%
\newcommand{\ccmark}{\color{black}{$\bigcircle$}}

\begin{document} 

\title{Multi-messenger constraints on the neutron-star equation of state and the Hubble constant}

\author{Tim Dietrich$^{1,2}$}
\author{Michael W.~Coughlin$^{3}$}
\author{Peter T.~H.~Pang$^{2,4}$}
\author{Mattia Bulla$^{5}$}
\author{Jack Heinzel$^{3,6,7}$}
\author{Lina Issa$^{5,8}$}
\author{Ingo Tews$^9$}
\author{Sarah Antier$^{10}$}

\affiliation{${}^1$ Institut f\"{u}r Physik und Astronomie, Universit\"{a}t Potsdam, 14476 Potsdam, Germany}
\affiliation{${}^2$ Nikhef, 1098 XG Amsterdam, The Netherlands}
\affiliation{${}^3$ School of Physics and Astronomy, University of Minnesota,
Minneapolis, MN 55455, USA}
\affiliation{${}^4$ Department of Physics, Utrecht University, 3584 CC Utrecht, The Netherlands}
\affiliation{${}^5$ Nordic Institute for Theoretical Physics (Nordita), 106 91 Stockholm, Sweden}
\affiliation{${}^6$ Department of Physics and Astronomy, Carleton College, Northfield, MN 55057, USA}
\affiliation{${}^7$ Artemis, Université Côte d’Azur, Centre National de la Recherche Scientifique, F-06304 Nice, France}
\affiliation{${}^8$ École normale supérieure, Universit\'e Paris-Saclay, 91190, Gif-sur-Yvette, France}
\affiliation{${}^9$ Theoretical Division, Los Alamos National Laboratory, Los Alamos, NM 87545, USA}
\affiliation{${}^{10}$ Astroparticule et Cosmologie, Universit\'e  de Paris, Centre National de la Recherche Scientifique,F-75013 Paris, France}

\date{18 Dec 2020}

\begin{abstract}
  Observations of neutron-star mergers based on distinct messengers, including gravitational waves and electromagnetic signals, can be used to study the behavior of matter denser than an atomic nucleus, and to measure the expansion rate of the Universe described by the Hubble constant.
  We perform a joint analysis of the gravitational-wave signal GW170817 with its electromagnetic counterparts AT2017gfo and GRB170817A, 
  and the gravitational-wave signal GW190425, both originating from neutron-star mergers. 
  We combine these with previous measurements of pulsars using X-ray and radio observations,
  and nuclear-theory computations using chiral effective field theory to constrain the neutron-star equation of state.
  We find that the radius of a $1.4$ solar mass neutron star is {$11.75^{+0.86}_{-0.81}\ \rm km$} at $90\%$ confidence and the Hubble constant is {$66.2^{+4.4}_{-4.2}\ \rm km \,Mpc^{-1}\, s^{-1}$} at $1\sigma$ uncertainty.
\end{abstract}

\maketitle


Multi-messenger observations of binary neutron-star (BNS) mergers, which employ different probes to observe the same astrophysical process, elucidate the properties of matter under extreme conditions and can be used to determine the expansion rate of the Universe described by the Hubble constant.
An example was the joint detection of gravitational waves (GWs), GW170817~\cite{TheLIGOScientific:2017qsa}, 
a gamma-ray burst (GRB), GRB170817A, 
a GRB afterglow arising from synchrotron radiation~\cite{Monitor:2017mdv}, and a kilonova, i.e., an electromagnetic (EM) signal in the optical, infrared, and ultraviolet bands originating from the radioactive decay of atomic nuclei created during a merger, AT2017gfo~\cite{GBM:2017lvd}, from the same astrophysical source. 
Using only GWs and the redshift of the host galaxy, this event led to an independent measurement of the Hubble constant~\cite{Abbott:2017xzu}.
It also placed constraints on the equation of state (EOS) of matter at densities higher than in the center of an atomic nucleus, e.g.,~\cite{Abbott:2018exr}.  
Moreover, GWs have been detected from another BNS merger, GW190425~\cite{Abbott:2020uma}, but no EM counterpart was observed~\cite{Coughlin:2019zqi}.
Joint observations of the mass and radius of the rapidly rotating neutron star (pulsar) PSR~J0030+0451 by the Neutron Star Interior Composition Explorer (NICER), e.g., \cite{Miller:2019cac}, have provided independent constraints on NS properties~\cite{Raaijmakers:2019qny}.
These build upon mass measurements of the pulsars PSR~J0740+6620~\cite{Cromartie:2019kug}, PSR~J0348+4042~\cite{Antoniadis:2013pzd}, and
PSR~J1614-2230~\cite{Arzoumanian:2017puf} using radio observations.

We combine the results from GW170817, GW190425, AT2017gfo, GRB170817A, PSR~J0030+0451, PSR~J0740+6620, PSR~J0348+4042, and
PSR~J1614-2230 with nuclear-theory calculations of the EOS, the latter using chiral effective field theory (EFT) predictions at low densities~\cite{materials}.
Previous studies have connected GW analyses to nuclear-physics predictions, e.g.,~\cite{Annala:2017llu,Capano:2019eae}, or performed Bayesian analyses of EM and GW signals, e.g.,~\cite{Coughlin:2018fis,Radice:2018ozg}, 
or combined GW and NICER results~\cite{Jiang:2019rcw,Raaijmakers:2019dks}. 
We combine all of these approaches, with the goal of providing improved constraints on the supranuclear EOS and measuring the Hubble constant.


\begin{figure*}[t]
    \centering
    \includegraphics[width=0.98\textwidth]{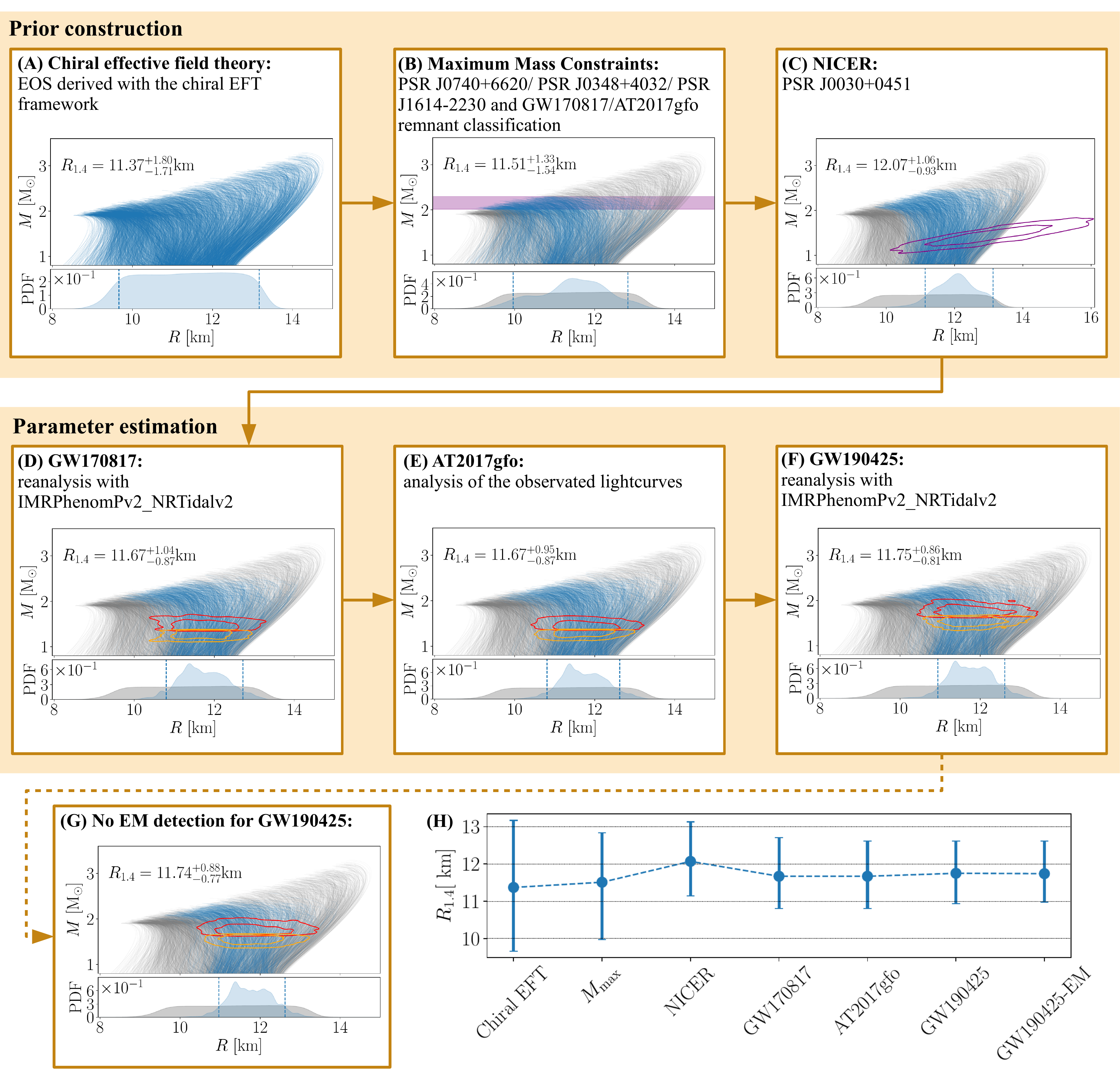}
    \caption{{\bf Multi-step procedure to constrain the neutron-star EOS.}
    In each panel, allowed (disallowed) EOSs are shown as blue (gray) lines. 
    Lower plots indicate the probability distribution function (PDF) for the radius of a 1.4 solar mass neutron star, with the 90\% confidence range indicated by dashed lines.
    (A) The set of EOSs from chiral EFT. 
    (B) The EOS set restricted by incorporating information from mass measurements of PSR~J0740+6620, PSR~J0348+4032, PSR~J1614-2230, and the maximum-mass constraints obtained from GW170817/AT2017gfo. 
    The 90\% confidence interval of the maximum mass posterior probability distribution is shown by a purple band.
    (C) The EOS set further restricted by the NICER mass-radius measurement of PSR~J0030+0451 (purple contours at 68\% and 95\% confidence).
    (D) Further restriction of the EOS set using Bayesian inference from our reanalysis of the GW170817 
    waveform. 
    Contours at 68\% and 95\% confidence show the mass-radius measurements of the primary (red) and secondary (orange) neutron stars.
    (E) We use the chirp mass, mass ratio, and the EOSs as Bayesian prior for our analysis of AT2017gfo. 
    (F) Further restrictions by analysing GW190425. This is our fiducial result.
    (G) Additional analysis assuming that GW190425 did not produce a detectable EM signal.
    (H) The radius constraint at each step of this analysis, with 90\% confidence ranges.}
    \label{fig:scheme}
\end{figure*}

We use a multi-step procedure, illustrated in Fig.~\ref{fig:scheme}, to incorporate 
constraints from nuclear theory and from astrophysical observations.
Our analysis begins with a newly-constructed set of $5000$ EOSs~\cite{materials} that provide possible descriptions of the structure of NSs (Fig.~\ref{fig:scheme}A). 
At low densities, these EOSs are constrained by microscopic calculations using chiral EFT interactions and computational many-body methods. Chiral EFT is a systematic theory for nuclear forces that describes the interactions in terms of nucleon and pion degrees of freedom and is consistent with the symmetries of quantum chromodynamics~\cite{Epelbaum:2008ga}. 
The resulting forces are arranged in an order-by-order expansion, which is then truncated at a certain level. 
This systematic scheme allows for the estimation of theoretical uncertainties from missing higher-order contributions to the nuclear interactions. 
The resulting nuclear Hamiltonians are inserted into the Schr\"odinger equation, which has been solved using quantum Monte Carlo methods~\cite{Carlson:2015}.
Chiral EFT might be valid up to $2n_{\rm sat}$~\cite{Tews:2018kmu}, where $n_{\rm sat}$ is the nuclear saturation density, 
$n_{\rm sat}=0.16$ fm$^{-3}$. 
Beyond that, chiral EFT interactions and their uncertainty estimates are not reliable.
We adopt a more conservative limit and constrain our EOSs with chiral EFT calculations up to densities of $1.5\, n_{\rm sat}$.
At densities above that limit, we employ a model-agnostic parametric expansion scheme that represents the EOS in the speed of sound plane~\cite{Tews:2018kmu} and ensures consistency with causality. 

We then restrict the set of EOSs by including astrophysical constraints. 
In a first step, we begin by enforcing a maximum NS mass $M_{\rm max}$ with an upper bound of $M_{\rm max} \leq 2.16^{+0.17}_{-0.15}$ solar masses ($M_\odot$) at $2\sigma$ uncertainty~\cite{Rezzolla:2017aly,materials}. 
This upper bound was derived by assuming that the final merger remnant of GW170817
was a black hole~\cite{Rezzolla:2017aly}. 
We derived a lower bound for the maximum mass by combining radio observations of
PSR~J0740+6620~\cite{Cromartie:2019kug}, PSR~J0348+4042~\cite{Antoniadis:2013pzd}, and
PSR~J1614-2230~\cite{Arzoumanian:2017puf}.
The resulting distribution for the maximum mass and the updated EOS set are shown in Fig.~\ref{fig:scheme}B.
For comparisons with other works, we calculate the radius of a typical 1.4$M_\odot$ NS at 90\% confidence. 
The corresponding radii at each stage of our analysis are shown in Fig.~\ref{fig:scheme}H.

\begin{table}[t]
\caption{\textbf{Comparison with selected radius constraints from multi-messenger observations.} 
For each reference, we indicate if chiral EFT input, constraints from heavy-pulsar mass measurements (Heavy PSRs), 
maximum-mass constraints obtained from GW170817/AT2017gfo ($M_{\rm max}$), 
GW constraints from GW170817 or GW190425, constraints from kilonova light curves (AT2017gfo),
constraints from the GRB afterglow (GRB170817A), and constraints from NICER have been used.  
We indicate with {\cmark} if either the full posterior probability distribution or a Bayesian Inference was employed, 
{\ccmark} if some information was included without performing a Bayesian analysis or including 
the full posterior probability distribution, and {\xmark} if the information was not included in the study.
Stated radius uncertainties represent 90\% confidence intervals, where for \cite{Radice:2018ozg} we also include systematic uncertainties as stated by the authors.}
\centering
\begin{tabular}{l|c|c|c|c|c|c|c|c|c}
     \begin{footnotesize} Reference \end{footnotesize}& 
     \begin{footnotesize} \rotatebox[origin=c]{82}{Chiral EFT} \end{footnotesize} & 
     \begin{footnotesize} \rotatebox[origin=c]{82}{Heavy PSRs} \end{footnotesize} &   
     \begin{footnotesize} \rotatebox[origin=c]{82}{$M_{\rm max}$ (remnant)} \end{footnotesize} &   
     \begin{footnotesize} \rotatebox[origin=c]{82}{GW1700817} \end{footnotesize}& 
     \begin{footnotesize} \rotatebox[origin=c]{82}{AT2017gfo} \end{footnotesize}& 
     \begin{footnotesize} \rotatebox[origin=c]{82}{GRB170817A} \end{footnotesize}& 
     \begin{footnotesize} \rotatebox[origin=c]{82}{NICER} \end{footnotesize}& 
     \begin{footnotesize} \rotatebox[origin=c]{82}{GW190425} \end{footnotesize} &    
     \begin{footnotesize} $R_{1.4M_\odot}$ \rm [km]\end{footnotesize}\\
        \hline
\begin{footnotesize} This work \end{footnotesize}
 &  yes  & \cmark & \cmark & \cmark & \cmark & \cmark & \cmark & \cmark & \begin{footnotesize} {$11.75^{+0.86}_{-0.81}$} \end{footnotesize} \\
        
\begin{footnotesize} \cite{Raaijmakers:2019dks}\end{footnotesize} & yes &  
\cmark & \xmark & \cmark & \xmark & \xmark & \cmark & \xmark & \begin{footnotesize}  $ [11.63, 13.26]$ \end{footnotesize} \\

\begin{footnotesize}  \cite{Capano:2019eae}\end{footnotesize} & yes &  
\ccmark & \ccmark & \cmark & \xmark & \xmark & \xmark & \xmark & \begin{footnotesize}  $11.0^{+0.9}_{-0.6}$ \end{footnotesize} \\

\begin{footnotesize} \cite{Coughlin:2018fis}\end{footnotesize} & no &  
\cmark & \ccmark & \cmark & \cmark & \ccmark & \xmark & \xmark & \begin{footnotesize}  $[11.3, 13.5]$ \end{footnotesize} \\

\begin{footnotesize}  \cite{Radice:2018ozg}\end{footnotesize} & no &  
\xmark & \xmark & \cmark & \ccmark & \xmark & \xmark & \xmark & \begin{footnotesize}  $(12.2^{+1.0}_{-0.8} \pm 0.2)$ \end{footnotesize} \\

\begin{footnotesize}  \cite{Abbott:2018exr}\end{footnotesize} & no &   
\ccmark & \xmark  & \cmark & \xmark & \xmark & \xmark & \xmark & $11.9^{+1.4}_{-1.4}$\\

\begin{footnotesize} 
 \cite{Annala:2017llu}\end{footnotesize} & yes  & 
\ccmark & \xmark & \ccmark & \xmark & \xmark & \xmark & \xmark &  \begin{footnotesize} $[9.9, 13.6]$\end{footnotesize}  \\

\end{tabular}
\label{tab:comparison}
\end{table} 

\begin{figure*}[t]
    \centering
    \includegraphics[width=0.56\textwidth]{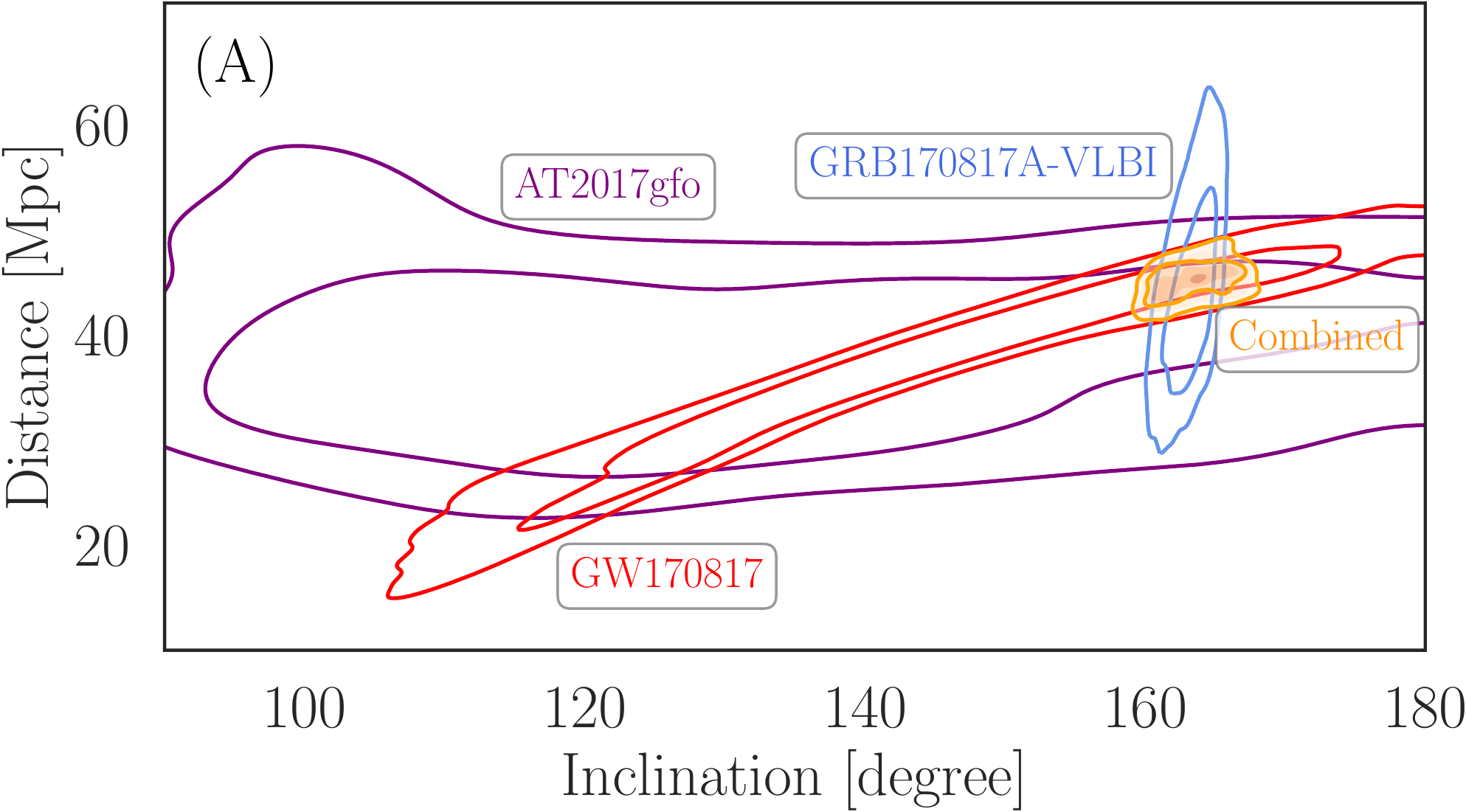}
    \hfill
    \includegraphics[width=0.43\textwidth]{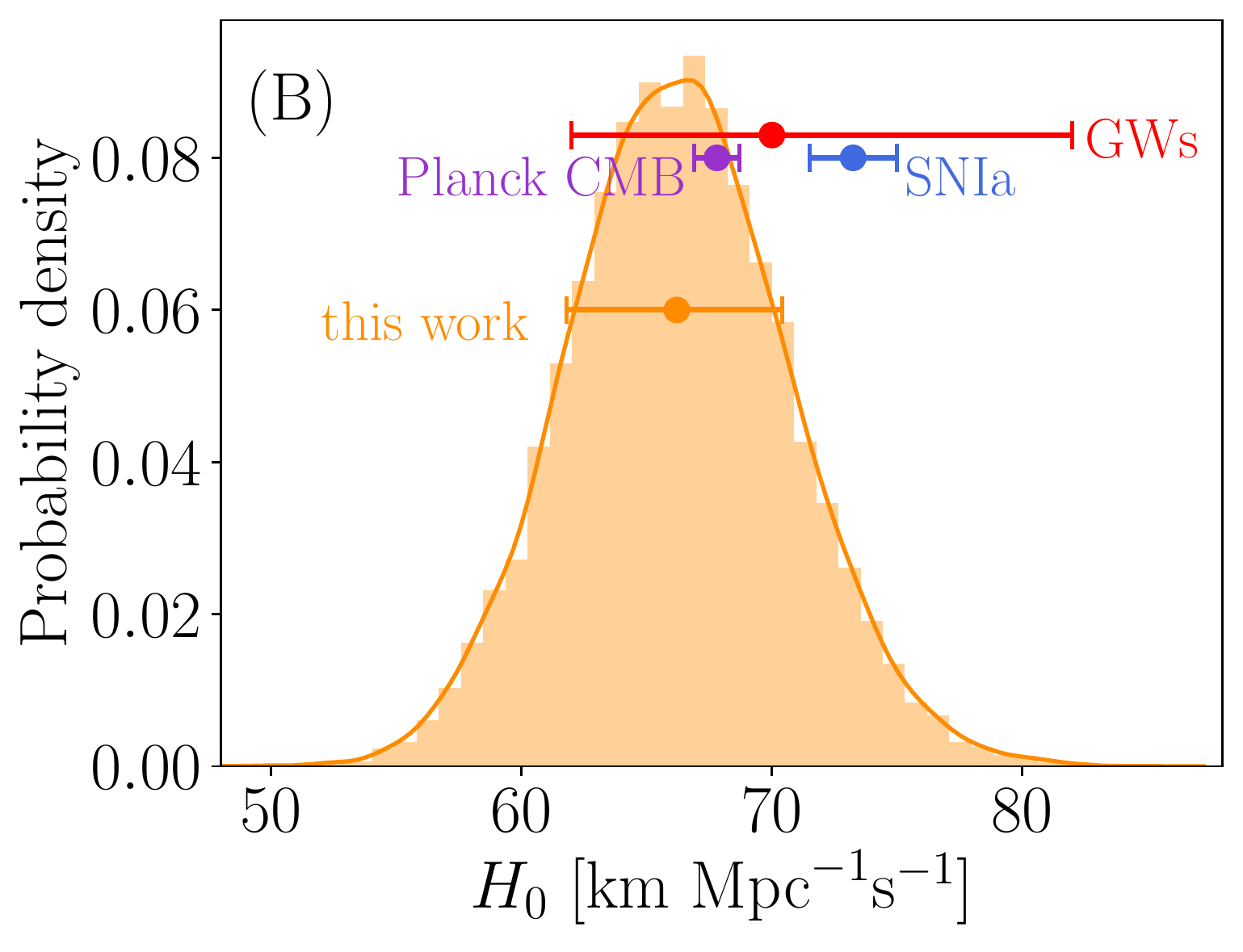}    
    \caption{\textbf{Distance-inclination constraints and Hubble constant measurement.}
    (A) Estimated distance and inclination of GW170817 from the GW waveform (red) and AT2017gfo analysis (purple) and the radio interferometry constraint~\cite{Hotokezaka:2018dfi} derived from GRB170817A (blue).
    The combined distance-inclination measurement is shown in orange.
    Contours are shown at 68\% and 95\% confidence.
    (B) Hubble constant estimate from our combined inclination measurement (orange histogram).
    Symbols mark the most probable values and $1\sigma$ uncertainties from this work (orange), the Planck measurement of the Cosmic Microwave Background~\cite{Ade:2015xua} (Planck CMB, purple), 
    the Hubble measurement via type-Ia supernovae~\cite{Riess:2016jrr} (SNIa, blue), 
    and the Hubble estimate from GW170817 alone~\cite{Abbott:2017xzu} (GWs, red).}
    \label{fig:D_i0}
\end{figure*}

In the next step, we include the NICER results~\cite{materials} using the joint posterior probability density function for mass and radius for the best fit model of Ref.~\cite{Miller:2019cac} shown in Fig.~\ref{fig:scheme}C.
We assign a probability to each EOS based on the maximum NS mass and NICER constraints.

By sampling over the obtained EOS set using their precomputed probabilities, we analyze GW170817~\cite{materials}, where 
NS properties are inferred from GW signals through tidal effects that are larger for NSs with smaller masses and larger radii. 
We employ the \textsc{parallel bilby} software~\cite{Smith:2019ucc} 
and the GW waveform model \texttt{IMRPhenomPv2\_NRTidalv2}~\cite{Dietrich:2019kaq}
for cross-correlation with the observed GW data~\cite{TheLIGOScientific:2017qsa}, inferring the binary properties from the measured signal. 
This model is an updated version of the waveform approximant 
\texttt{IMRPhenomPv2\_NRTidal} which has been used in previous analyses of GW170817~\cite{Abbott:2018wiz} and 
GW190425~\cite{Abbott:2020uma}. 

In the fourth step, we add constraints from AT2017gfo using a published light curve model~\cite{materials,Bulla:2019muo}. 
We use a Gaussian-Process-Regression framework to compute generic light curves 
for various ejecta-mass properties.
To connect the individual ejecta parameters to the properties of the system, 
we assume that the total ejecta mass $M_{\rm ej}$ 
is a sum of multiple components: dynamical ejecta $M^{\rm dyn}_{\rm ej}$, the material released during the merger process via shocks and torque, and disk-wind ejecta $\zeta M_{\rm disk}$:
$M_{\rm ej} = M^{\rm dyn}_{\rm ej} + \zeta M_{\rm disk} + \alpha$. 
The parameters $\alpha$, corresponding to a potentially unmodelled ejecta component, 
and $\zeta$, determining how much mass of the disk is ejected, 
are unknown free parameters. 
Our treatment of the dynamical ejecta follows previous work~\cite{Coughlin:2018fis}. 
Existing disk-wind ejecta models are known to be inappropriate for systems with high mass ratios. 
To overcome this issue, we include an explicit mass-ratio dependence in the disk-mass prediction~\cite{materials}.
The GW results for the chirp mass $\mathcal{M}_c = (m_1 m_2)^{3/5}(m_1 + m_2)^{-1/5}$, with $m_1$ and $m_2$ being the masses of the heavier and lighter NS, respectively, the mass ratio $q=m_1 m_2^{-1}$, and the EOS are used as priors for our analysis of AT2017gfo. 
This further constrains the EOS models (Fig.~\ref{fig:scheme}D).
Including all steps so far, we obtain the radius of a 1.4$M_\odot$ NS of {$R_{1.4M_\odot}=11.67^{+0.95}_{-0.87}\ \rm km$} at 90\% confidence.

These results can be further constrained by combining them with another observed BNS merger, GW190425~\cite{Abbott:2020uma}.  
Due to the high total mass of GW190425 of $3.4_{-0.1}^{+0.3} M_{\odot}$ at 90\% confidence, which suppresses tidal effects, we find that the inclusion of GW190425 does not improve the precision, but does slightly shift the median value within the uncertainty.
Our final estimate on the radius of a $1.4M_\odot$ NS is {$R_{1.4M_\odot}=11.75^{+0.86}_{-0.81}\ \rm km$} with 90\% confidence. 
We also explore an alternative ordering of individual analysis steps (Fig.~\ref{fig:analysis_interchange}) and systematic uncertainties due to the use of different GW models (Fig.~\ref{fig:systematics}), but obtain a consistent radius constraint (see Supplementary Text).

Several independent EM searches for counterparts to GW190425 observed large fractions of the possible sky area~\cite{Coughlin:2019zqi} (see Supplementary Text), suggesting that most of the appropriate region was searched but no EM signal was detected. 
To include this non-detection, we employ the same kilonova analysis as for GW170817, combining it with upper limits reported by the optical EM counterpart searches.
Using the distance information from the GW data, $159_{-71}^{+69}\, \rm Mpc$ at 90\% confidence level~\cite{Abbott:2020uma}, we obtain limits on the absolute magnitude of a potential counterpart. 
Using our light curve models, we rule out parts of the parameter space for which the predicted absolute magnitude would be above the obtained limit. 
Following this procedure, we arrive at a radius estimate of {$R_{1.4M_\odot}=11.74^{+0.88}_{-0.77}\ \rm km$} (90\% confidence) under the assumption that if GW190425 produced a detectable signal, it would have been found.
To be conservative, we omit this step from the subsequent analysis.

Our study includes information from GW170817, AT2017gfo, GRB170817A, GW190425, the NICER observation of PSR~J0030+0451, and the radio observations of PSR~J0740+6620, PSR~J0348+4032, PSR~J1614-2230.
Our approach allows for strong phase transitions in the EOS, the combination of multiple events, and the incorporation of EM non-detections. 
We compare our final result of {$R_{1.4M_\odot}=11.75^{+0.86}_{-0.81}\ \rm km$} with a selection of previous studies in Tab.~\ref{tab:comparison}.  
We note that the inclusion of additional astrophysical observations does not necessarily lead to tighter constraints (Fig.~\ref{fig:scheme}H) as
(i) the full combined posterior probability distributions are incorporated in the analysis and 
(ii) the number of events detected with multiple messenger remains very small.

In addition to EOS studies, we perform a measurement of the Hubble constant~\cite{materials}. 
For this purpose, we assume that measurable properties related to the kilonova, e.g., time-scale and color evolution of the ejecta, are connected to its intrinsic luminosity.
Theoretical kilonova predictions can be used to standardize kilonovae light curves and thereby measure their distances~\cite{Coughlin:2019vtv}. 
Combining the distance measurement with the redshift $z$ of the host galaxy NGC~4993, $z = 0.009783 \pm 0.000023$, constrains the Hubble constant~\cite{Abbott:2017xzu}.
We combine the distance and inclination measurements of the GW and kilonova analyses with the measurement using radio observations of the GRB afterglow  (Fig.~\ref{fig:D_i0})~\cite{materials,Hotokezaka:2018dfi}.
The comparison of a kilonova observation to a light curve model permits a large parameter range, due to the complexity of the model. 
Adopting two other kilonova models (see Supplementary Text) indicates that our kilonova constraints are conservative, but we note that it is not possible to test the robustness of different kilonova models with only one well-sampled kilonova observation (AT2017gfo).
Combining all these measurements leads to an improved distance constraint and an estimate of the Hubble constant of
{$H_0=66.2^{+4.4}_{-4.2}\ \rm km \,Mpc^{-1}\, s^{-1}$} at $1\sigma$ uncertainty (Fig.~\ref{fig:D_i0}).
We find that the radio inclination measurement reduces the existing uncertainty on the Hubble constant by more than the kilonova measurement, at least for this single event.
The uncertainty does not allow us to resolve the tension between measurements via type-Ia supernovae~\cite{Riess:2016jrr}
and the Planck measurement of the Cosmic Microwave Background~\cite{Ade:2015xua}, 
but our results indicate a preference for the latter and disfavor the measurement via type-Ia supernovae~\cite{Riess:2016jrr}. 

%
\bibliography{NMMA_sci.bbl}

{\bf Acknowledgments}:
We thank Kenta Hotokezaka for providing the posterior probability distribution samples of~\cite{Hotokezaka:2018dfi}. 
We are also grateful to Zoheyr Doctor, Reed Essick, and the anonymous referees for helpful comments on the manuscript.
{\bf Funding:}
T.D. acknowledges support by the European Union’s Horizon 2020 research and innovation program under grant agreement No 749145, BNSmergers. 
M.W.C. acknowledges support from the National Science Foundation with grant number PHY-2010970.
P.T.H.P. is supported by the research program of the Netherlands Organization for Scientific Research (NWO).
J.H. acknowledges support from the National Science Foundation with REU grant number NSF1757388.
I.T. is supported by the U.S. Department of Energy, Office of Science, Office of Nuclear Physics, under contract No.~DE-AC52-06NA25396, by the Laboratory Directed Research and Development program of Los Alamos National Laboratory under project number 20190617PRD1, and by the U.S. Department of Energy, Office of Science, Office of Advanced Scientific Computing Research, Scientific Discovery through Advanced Computing (SciDAC) program.
S.A. is supported by the CNES Postdoctoral Fellowship at Laboratoire Astroparticle et Cosmologie.
Computations were performed on the Minerva HPC cluster of the Max-Planck-Institute for Gravitational Physics, on SuperMUC-NG (LRZ) under project number pn56zo, and on HAWK (HLRS) under project number 44189.  
Computational resources were also provided by the Los Alamos National Laboratory Institutional Computing Program, which is supported by the U.S. Department of Energy National Nuclear Security Administration under Contract No.~89233218CNA000001, and by the National Energy Research Scientific Computing Center (NERSC), which is supported by the U.S. Department of Energy, Office of Science, under contract No.~DE-AC02-05CH11231.
{\bf Author contributions:}
Conceptualization: T.D., M.W.C., M.B., I.T.
Methodology: T.D., M.W.C., P.T.H.P., M.B., J.H., L.I., I.T., S.A.
Software: T.D., M.W.C., P.T.H.P., M.B., J.H., L.I.,  I.T.
Validation: T.D., M.W.C., P.T.H.P., M.B., I.T.
Formal analysis: T.D., M.W.C., P.T.H.P., I.T.
Resources: T.D., M.W.C., I.T.
Data curation: T.D., M.W.C., P.T.H.P., J.H., I.T.
Writing—original draft: T.D., M.W.C., P.T.H.P., M.B., I.T.
Writing—review and editing: T.D., M.W.C., P.T.H.P., J.H., L.I., I.T., S.A.
Visualization: T.D., M.W.C., P.T.H.P. 
Supervision: T.D., M.W.C., I.T.
Project administration: T.D., M.W.C., I.T.
Funding acquisition: T.D., M.W.C., I.T. 
{\bf Competing interests:}
The authors declare no competing interests.
{\bf Data and Materials availability:}
All data are available in the manuscript or the Supplementary Materials. 
Full posterior data samples of our analysis and code patches to repeat the study can be downloaded from~\cite{dietrich_tim_2020_4114141} and from \url{https://github.com/diettim/NMMA/}.
All employed GW models are implemented in the publicly available software \textsc{LALSuite} at \url{https://git.ligo.org/lscsoft}. 
The \textsc{bilby} and \textsc{parallel bilby} softwares are available at \url{https://git.ligo.org/lscsoft/bilby} and \url{https://git.ligo.org/lscsoft/parallel_bilby}, respectively. 
The \textsc{gwemlightcurve} software is available at \url{https://gwemlightcurves.github.io/}.
The exact code versions of \textsc{bilby}, \textsc{parallel bilby}, \textsc{LALSuite}, and \textsc{gwemlightcurve} that we have employed for this work are also available at~\cite{dietrich_tim_2020_4114141}.  
 The gravitational wave data that we have analysed in this work was obtained from the Gravitational Wave Open Science Center (GWOSC) at \url{https://www.gw-openscience.org} and the NICER data are taken from~\texttt{doi:10.5281/zenodo.3473466}. \\ 


\clearpage
  
\newpage  
\onecolumngrid
  
\renewcommand\thefigure{S\arabic{figure}} 
\setcounter{figure}{0}
\renewcommand\thetable{S\arabic{table}} 
\setcounter{table}{0}
\renewcommand\theequation{S\arabic{equation}} 
\renewcommand\thepage{S\arabic{page}}
\setcounter{page}{1}

\begin{center}
\vspace{0.7cm}

{\Large{\bf Supplementary Materials for}}
\vspace{0.3cm}

{\large{Multi-messenger constraints on the neutron-star equation of state and the Hubble constant}}
\vspace{0.3cm}

Tim Dietrich, Michael W.~Coughlin, Peter T.~H.~Pang, Mattia Bulla, Jack Heinzel, Lina Issa, Ingo Tews, Sarah Antier\\
\vspace{0.3cm}

Correspondence to: tim.dietrich@uni-potsdam.de
\end{center}

\section*{Materials and Methods}

\subsection*{Chiral effective field theory and the neutron-star equation of state}

Microscopic nuclear interactions are governed by multiple processes, e.g., various longer-range meson exchanges between two or more nucleons or short-range processes that are typically modeled by contact interactions. 
Nuclear effective field theories, like chiral EFT~\cite{Weinberg1990,Weinberg1991,vanKolck:1994yi,Epelbaum:2008ga,Machleidt:2011zz}, provide a framework for arranging the large number of operator structures for nuclear interactions.
\\
Nuclear EFTs start from the most general Lagrangian that is consistent with all symmetries of the fundamental theory of strong interactions, quantum chromodynamics, and that describes the various interaction mechanisms. 
In chiral EFT, this Lagrangian is written in terms of nucleon and pion degrees of freedom, and includes pion-exchange interactions as well as nucleon-contact interactions~\cite{Weinberg1990,Weinberg1991}. 
The latter absorb short-range effects, e.g., exchanges of heavier mesons, and depend on coupling constants that have to be adjusted to experimental data. 
Because this Lagrangian contains an infinite number of terms, it is then expanded in powers of momenta $p$ over the breakdown scale $\Lambda_b$. 
In addition to two-nucleon interactions, the chiral EFT expansion includes many-body forces, where three or more nucleons interact with each other. 
This results in a systematic and consistent expansion of two- and many-body nuclear forces, which can be truncated at a chosen order.
By going to higher orders in the expansion, nuclear interactions can be systematically improved.
By calculating results order-by-order,
theoretical uncertainties due to our incomplete understanding of nuclear interactions can be quantified~\cite{Epelbaum:2015epja}.
\\
Chiral interactions allow an extrapolation of nuclear interactions away from experimentally accessible systems to those that are difficult or impossible to measure in terrestrial laboratories, e.g., the neutron-rich matter in the core of NSs. 
However, chiral interactions are limited to momenta $p<\Lambda_b\approx 600$ MeV~\cite{Melendez:2017phj}. At larger momenta, chiral interactions are not reliable because short-range (high-energy) physics that was absorbed by the coupling constants needs to be explicitly included. 

The EOSs used in this work are constrained at low densities by quantum Monte Carlo calculations of neutron matter~\cite{Carlson:2015,Lynn:2019rdt} at temperature $T=0$, using the auxiliary field diffusion Monte Carlo approach and chiral EFT interactions in their local formulation~\cite{Gezerlis:2013,Gezerlis:2014,Lynn:2016}.
The unknown coupling constants in chiral EFT are determined by fitting the nuclear Hamiltonians order-by-order to experimental data~\cite{Tews:2020hgp}. 
The interactions used here were fitted to two-nucleon scattering data, the $^4$He ground state energy, and neutron-$\alpha$ scattering phase shifts~\cite{Gezerlis:2014,Lynn:2016}. 
The order-by-order convergence of this approach remains valid up to densities of twice the nuclear saturation density~\cite{Tews:2018kmu}. 
To be more conservative, we employ these calculations up to densities of $1.5\, n_{\rm sat}$ to constrain the NS EOS below that density. 
First, we extend the results to matter in $\beta$-equilibrium and add a crust~\cite{Tews:2016ofv}. 
Then, we extend our EOS models to densities beyond $1.5\, n_{\rm sat}$ by employing a model-agnostic parametric expansion scheme that represents the EOS in the speed of sound plane~\cite{Capano:2019eae,Tews2018,Tews:2019cap,Greif:2018njt}. 
For each EOS, we sample a set of six randomly distributed points in the speed of sound plane at baryon densities between $1.5\, n_{\rm sat}$ and $12\, n_{\rm sat}$ and connect them by line segments. 
We found that NS properties are not very sensitive to the number of line segments when varying it between 5-10. 
This construction by design remains causal and stable at all densities, $0\leq c_S \leq c$, with the speed of sound $c_S$ and the speed of light $c$.  
From the speed-of-sound curves, we reconstruct the EOSs and solve the Tolman-Oppenheimer-Volkoff (TOV) equations~\cite{TOV,TOV2} to extract NS structure properties. 
For each sampled EOS, we construct a second EOS that includes a segment with $c_S=0$ with random onset density and width, to simulate EOSs with strong first-order phase transitions.
We sampled 5000 different EOSs to produce a uniform prior on the radius of a typical $1.4 M_{\odot}$ NS (Fig.~\ref{fig:scheme}A).

Similar to commonly used polytropic expansion schemes~\cite{Hebeler:2013nza}, the speed-of-sound extension does not make any assumptions about degrees of freedom at higher densities, and includes many possible density dependencies for the EOS at high densities.
For example, this extension includes regions of sudden stiffening or sudden softening, as would be expected from a strong first-order phase transition. 

\subsection*{Incorporation of the maximum mass neutron-star constraints}

For the inclusion of the astronomical constraints on the EOSs, we adopt a Bayesian approach~\cite{Alvarez-Castillo:2016oln,Miller:2019nzo}, and express the constraints in terms of likelihood functions that can be used for the GW and EM analysis. 

\begin{figure}[h!]
    \centering
    \includegraphics[width=0.5\textwidth]{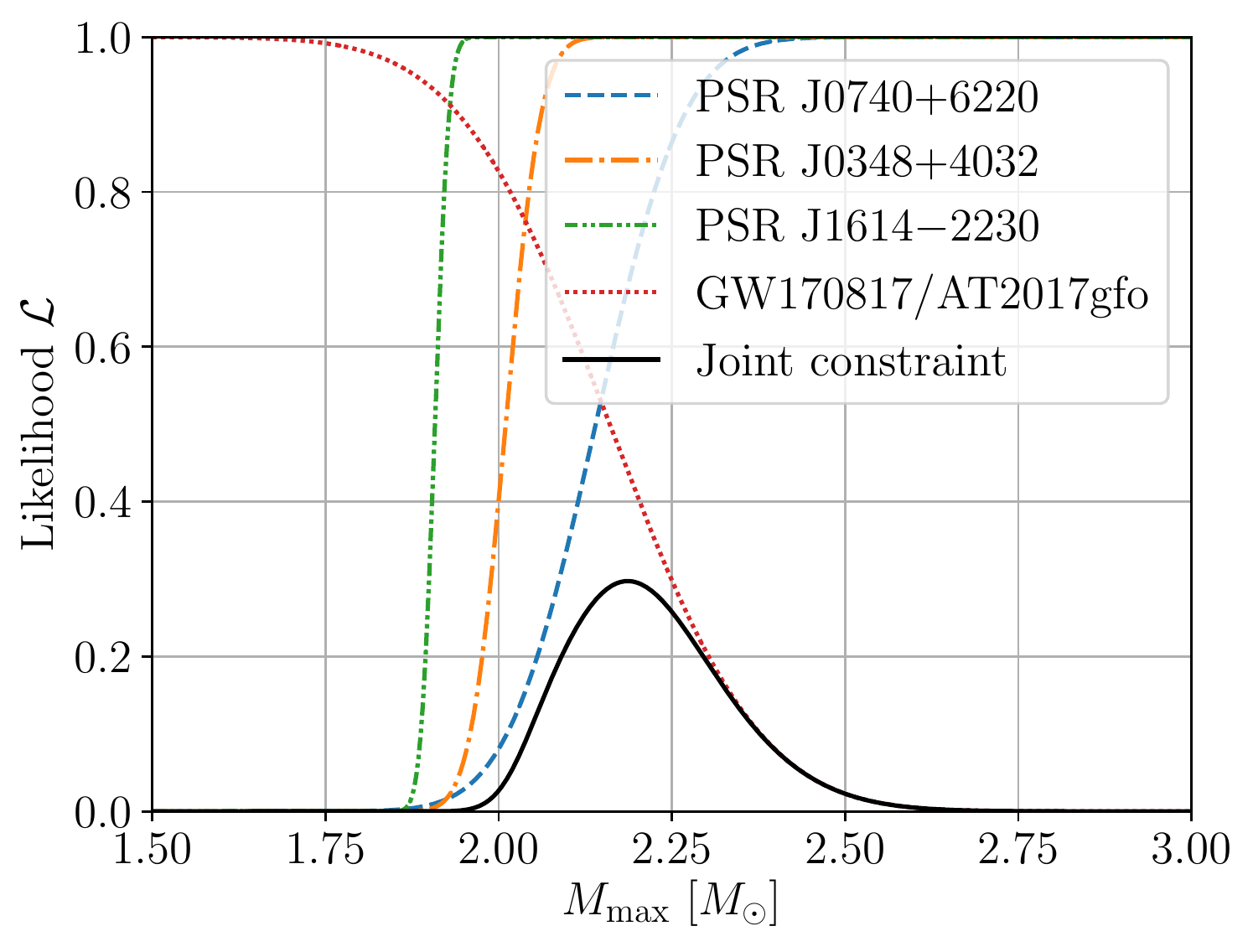}
	\caption{\textbf{Combined likelihood of the maximum mass.} Shown are the constraints from radio observations of 
	PSR~J0740+6620~\cite{Cromartie:2019kug}, PSR~J0348+4032~\cite{Antoniadis:2013pzd}, 
	and PSR~J1614-2230~\cite{Arzoumanian:2017puf} (lower bounds) 
	from the remnant classification of GW170817/AT2017gfo as a black hole~\cite{Rezzolla:2017aly} (upper bound),
	and the joint constraint (black line).
	}
    \label{fig:MTOV_likelihood}
\end{figure}

We have used constraints on the lower bound of the maximum NS mass $M_{\textrm{max}}$ given by the mass measurements of pulsars PSR~J0740+6620~\cite{Cromartie:2019kug}, PSR~J0348+4032~\cite{Antoniadis:2013pzd},
PSR~J1614-2230~\cite{Arzoumanian:2017puf}, and a constraint on the upper bound on $M_{\textrm{max}}$~\cite{Rezzolla:2017aly} of $M_{\textrm{max}}~=~2.16^{+0.17}_{-0.15}M_{\odot}$ at $95$\% confidence. 
Similar upper bounds on $M_{\textrm{max}}$ have also been obtained in different studies, e.g., 
$M_{\rm max}\lesssim 2.17M_\odot$ at $90$\% confidence~\cite{Margalit:2017dij},
$M_{\rm max} \lesssim 2.3M_{\odot}$~\cite{Shibata:2019ctb}, 
or $M_{\rm max} \lesssim 2.16-2.28 M_\odot$~\cite{Ruiz:2017due}.
The corresponding likelihood $\mathcal{L}_{\rm{M_{max}}}$ is given by
\begin{equation}
\begin{aligned}
	\mathcal{L}_{\rm{M_{max}}}(\textrm{EOS}) &= \mathcal{L}_{\rm{M_{max}}}(M_{\textrm{max}})\\
	&= \prod_i \textrm{CDF}(M_{\textrm{max}}, \mathcal{N}(M^{\textrm{PSR}}_{i}, \sigma^{\textrm{PSR}}_{i}))\\
	&\times (1-\textrm{CDF}(M_{\textrm{max}}, \mathcal{N}(2.16M_{\odot}, 0.17M_{\odot}))),
\end{aligned}
\end{equation}
where $\textrm{CDF}(x, \mathcal{N}(\mu, \sigma))$ is the cumulative distribution function corresponding to a normal distribution $\mathcal{N}(\mu, \sigma)$ evaluated at $x$. 
$M^{\textrm{PSR}}_{i}$ and $\sigma^{\textrm{PSR}}_{i}$ are the mass measurement and the $1$-$\sigma$ uncertainty reported for the pulsars that we included for the analysis, respectively. 
The values for $M^{\textrm{PSR}}_{i}$ and $\sigma^{\textrm{PSR}}_{i}$ are tabulated in Tab.~\ref{tab:PSR_mass}.
For the upper bound on $M_{\textrm{max}}$, we take a more conservative uncertainty, adopting the $95\%$ credible range as the standard deviation for the likelihood input. 
In the likelihood $\mathcal{L}_{\rm{M_{max}}}$, we have approximated the measurements~\cite{Cromartie:2019kug,Antoniadis:2013pzd,Arzoumanian:2017puf} and estimates~\cite{Rezzolla:2017aly} as Gaussian.
The final likelihood is shown in Fig.~\ref{fig:MTOV_likelihood}.

\begin{table}[h!]
\centering
\caption{\textbf{Summary of the heavy-pulsar mass measurements.} The masses $M^{\rm PSR}$ and their $1$-$\sigma$ uncertainties $\sigma^{\rm PSR}$ reported for the pulsars included in this analysis.}
\begin{tabular}{l|l|l|l}
        Pulsar & $M^{\textrm{PSR}}$ $[M_{\odot}]$ & $\sigma^{\textrm{PSR}}$ $[M_{\odot}]$ & Reference\\
        \hline
	PSR~J0740+6620 & 2.14 & 0.1 & \cite{Cromartie:2019kug} \\
	PSR~J0348+4032 & 2.01 & 0.04 & \cite{Antoniadis:2013pzd}\\
	PSR~J1614-2230  & 1.908 & 0.016 & \cite{Arzoumanian:2017puf}\\
\end{tabular}
\label{tab:PSR_mass}
\end{table}

\subsection*{Coherent incorporation of NICER data}

\begin{figure}[h!]
    \centering
    \includegraphics[width=0.55\textwidth]{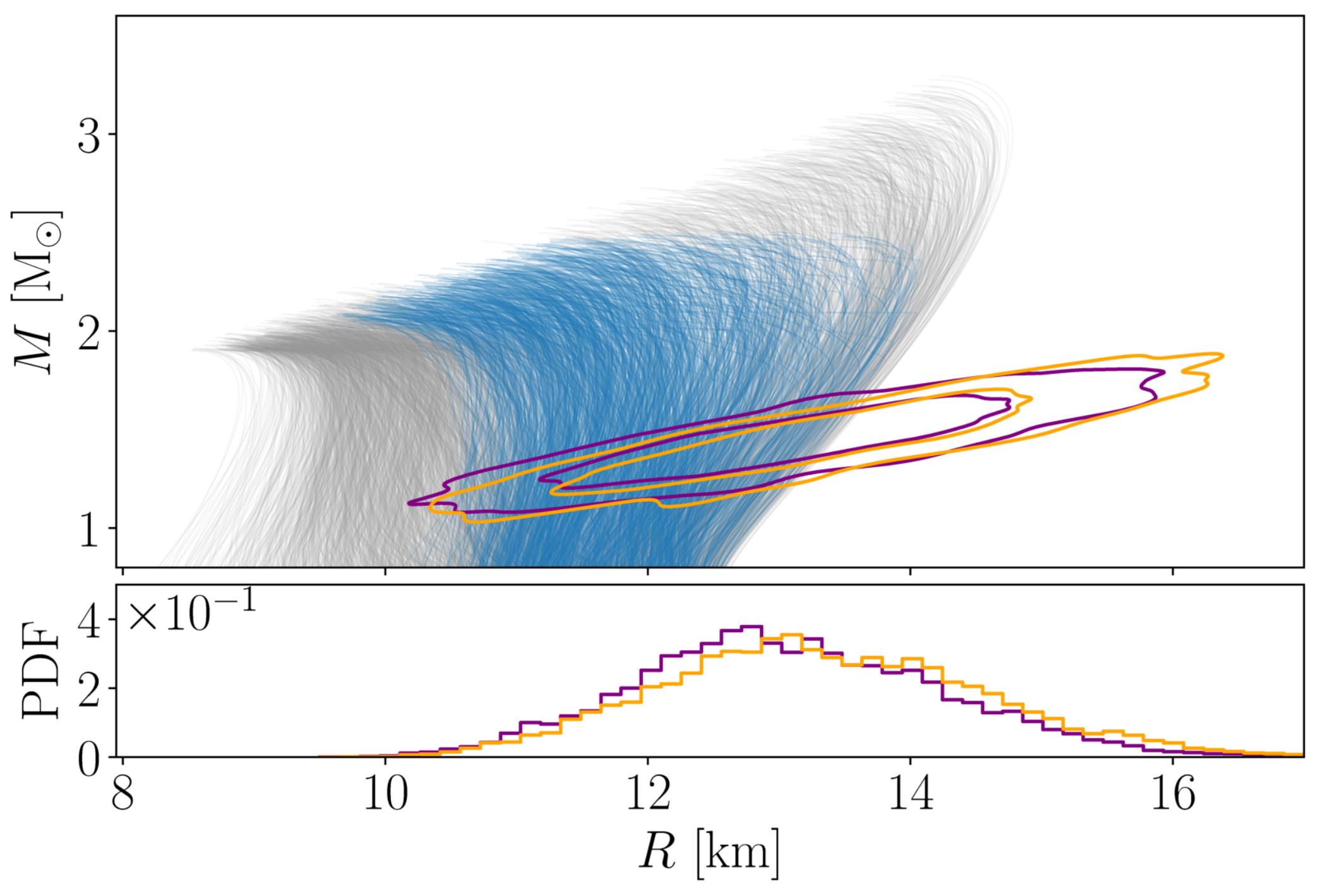}
	\caption{\textbf{
	Comparison of different NICER analysis models}. Shown are the 2D posterior probability distributions for the mass and radius of PSR J0030+0451 inferred with a three-oval spot model (purple) and a two-oval spot model (orange)~\cite{Miller:2019cac} in comparison with our EOS constraint at this analysis step, cf. Fig.~\ref{fig:scheme}. 
	Contours are shown at 68\% and  95\%. }
    \label{fig:NICER_models}
\end{figure}

For the NICER data~\cite{Bogdanov:2019qjb}, we use the results from Ref.~\cite{Miller:2019cac} 
where a Bayesian inference approach was used to analyze the energy-dependent thermal
X-ray waveform of PSR~J0030+0451.
We employ the samples obtained with a three-oval, uniform-temperature spots model~\cite{Miller:2019cac,miller_m_c_2019_3473466}. 
This model provides agreement with the observed NICER data and constrains the mass and radius of PSR~J0030+0451
to be $M=1.44^{+0.15}_{-0.14} M_\odot$ and 
$R = 13.02^{+1.24}_{-1.06} \rm km$ (both at $1\sigma$ uncertainty). 
The inferred mass-radius posterior probability distributions are not dominated by systematic uncertainties and inferred parameters are in agreement for different models~\cite{Miller:2019cac,Riley:2019yda}; as a comparison, the results 
for the two-oval spot model are shown together with the three-oval spots model in Fig.~\ref{fig:NICER_models}. 
Further comparisons can be found in Refs.~\cite{Miller:2019cac,Riley:2019yda}.

The corresponding likelihood $\mathcal{L}_{\textrm{NICER}}$ is given by
\begin{equation}
\begin{aligned}
	\mathcal{L}_{\textrm{NICER}}(\textrm{EOS}) &= \int d\!M d\!R\ p_{\textrm{NICER}}(M, R)\pi(M, R |\textrm{EOS})\\
	&= \int d\!M d\!R\ p_{\textrm{NICER}}(M, R)\delta(R-R(M,\textrm{EOS}))\\
	&= \int d\!M\ p_{\textrm{NICER}}(M, R=R(M, \textrm{EOS})),
\end{aligned}
\end{equation}
where $p_{\textrm{NICER}}(M, R)$ is the joint-posterior probability distribution of mass and radius of PSR~J0030+0451 as measured by NICER and we use the fact that the radius is a function of mass for a given EOS.

The joint-constraint likelihood $\mathcal{L}_{\textrm{Joint}}$ combining the 
maximum mass and the NICER information is given by
\begin{equation}
	\mathcal{L}_{\textrm{Joint}}(\textrm{EOS}) = \mathcal{L}_{\textrm{NICER}}(\textrm{EOS}) \times \mathcal{L}_{\rm{M_{max}}}(\textrm{EOS}).
\end{equation}
$\mathcal{L}_{\textrm{Joint}}(\textrm{EOS})$ is then taken as an input for our further analysis of GW170817, AT2017gfo, and GW190425.

\subsection*{Gravitational-Wave Analysis}
\label{sec:GW}

We use the \textsc{bilby} software~\cite{Ashton:2018jfp}
to reanalyze the observational data for GW170817~\cite{GW170817_open_data} and 
GW190425~\cite{GW190425_open_data}.
We ran \textsc{parallel bilby}~\cite{Smith:2019ucc} on $800$ cores to obtain posterior probability distributions within a few hours on the high-performance computing (HPC) clusters Minerva at the Max-Planck-Institute for Gravitational Physics, on SuperMUC-NG at the Leibniz Supercomputing Centre, or on the HAWK cluster of the High-Performance Computing Center Stuttgart.  
The GW signals are analysed within a frequency interval $f\in [23, 2048\rm] Hz$ which covers the full inspiral of the BNS coalescence. 
Frequency-dependent spline calibration envelopes~\cite{calibration_envelope} are introduced 
into the waveform templates to counteract the potential systematics due to the uncertainties in 
the detectors' calibrations~\cite{PhysRevD.96.102001,Viets_2018}.
We adopt the power spectral density estimated with \textsc{BayesWave}~\cite{Cornish_2015,PhysRevD.91.084034}. 
For our analysis, we employ the \texttt{IMRPhenomPv2\_NRTidalv2} (\texttt{NRTidalv2}) waveform model~\cite{Dietrich:2019kaq}. 

\subsection*{AT2017gfo}
\label{sec:kilonova}

\subsubsection*{Kilonova modelling}

For the assessment of systematic uncertainties, 
we compare multiple light curve models~\cite{Bulla:2019muo,Kasen:2017sxr}.

\begin{figure}[h!]
    \centering
    \includegraphics[width=0.50\textwidth]{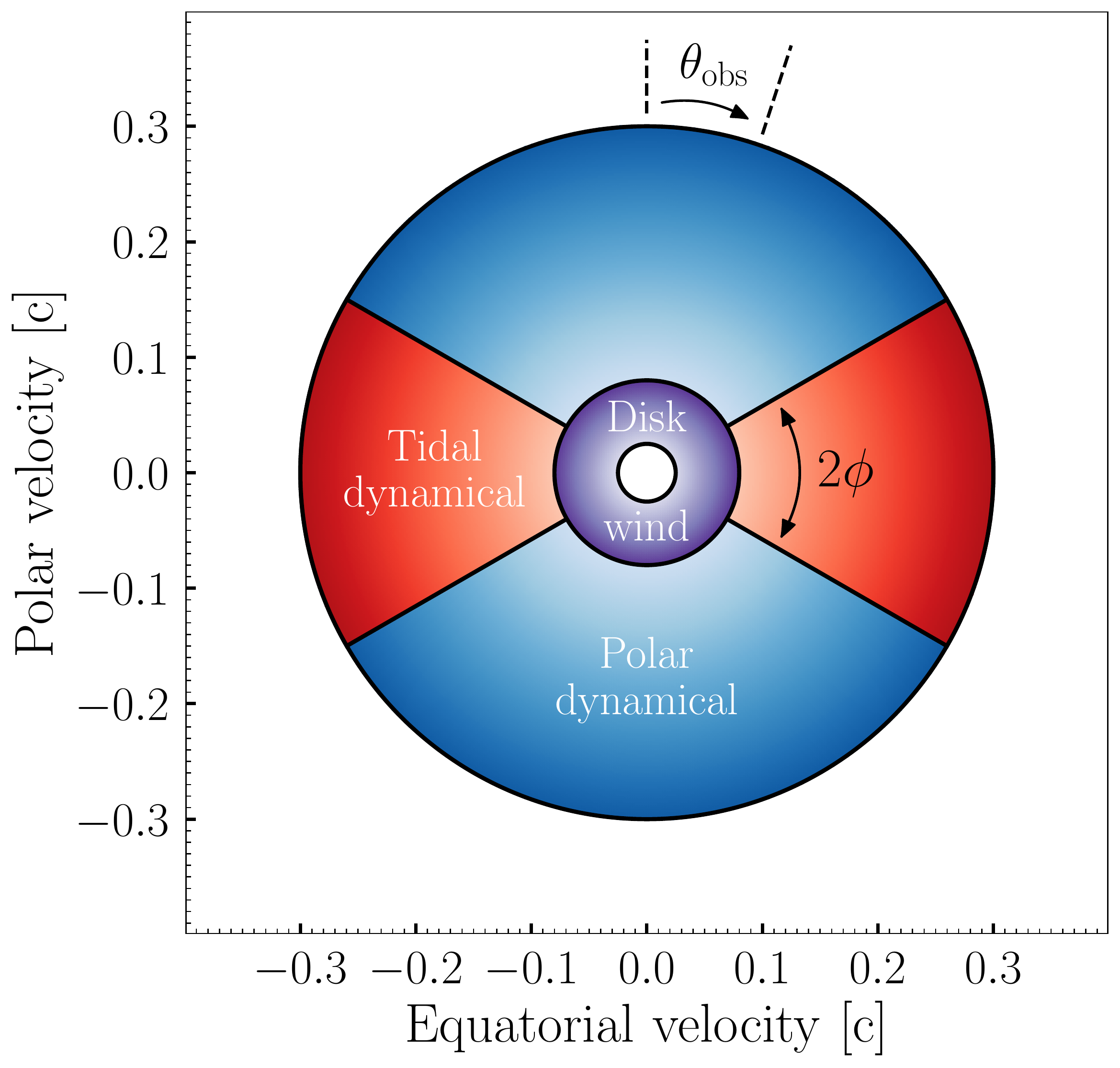}
    \caption{\textbf{Geometry employed in the kilonova description of Model I.} Different colors refer to the different lanthanide fractions of the individual ejecta components: tidal dynamical (red), polar dynamical (blue), and disk wind (purple).}
    \label{fig:Bulla_geometry}
\end{figure}

\textit{Model I (standard model):} This model uses Spectral Energy Distributions (SEDs) simulated using 
the multi-dimensional Monte Carlo radiative transfer code \textsc{possis}~\cite{Bulla:2019muo}. 
We use a model grid with modifications to the underlying physics and the assumed geometry for the ejecta. 
Compared to previous work~\cite{Bulla:2019muo}, we introduce two changes to the physics: thermalization efficiencies are taken from Ref.~\cite{Barnes:2016umi} 
and the temperature is estimated in each grid cell and at each time from the mean intensity of the radiation field 
(inferred from the density and local energy deposition from radioactive decay). 
In terms of the adopted geometry, we run calculations for geometries similar to, e.g., Refs.~\cite{Dietrich:2016fpt,Perego:2017wtu,Kawaguchi:2019}, see Fig.~\ref{fig:Bulla_geometry}, which were obtained from numerical relativity simulations. 
A first component represents the dynamical ejecta, which have velocities ranging from
the minimum velocity of the dynamical ejecta \mbox{$v_\mathrm{min}^\mathrm{dyn}=0.08$\,c} to 
the maximum velocity of the dynamical ejecta \mbox{$v_\mathrm{max}^\mathrm{dyn}=0.3$\,c}, 
are characterised by an ejecta mass $M_\mathrm{ej}^\mathrm{dyn}$, and have a 
lanthanide-rich composition within an angle $\pm\Phi$ about the equatorial 
plane and a lanthanide-free composition otherwise.
The dynamical ejecta correspond to a high-velocity portion of the geometry adopted in Ref.~\cite{Bulla:2019muo}. 
The main source of opacity in kilonova ejecta is given by bound-bound line transitions, in which electrons move between two bound states of atoms or ions. The bound-bound opacities $\kappa_\mathrm{bb}$ assumed for the dynamical ejecta are wavelength- and time-dependent, reaching values of $\kappa_\mathrm{bb}=1$\,cm$^2$ g$^{-1}$ at $1\mathrm{\upmu m}$ and 1.5\,d for the lanthanide-rich and $\kappa_\mathrm{bb}=5\times10^{-3}$\,cm$^2$ g$^{-1}$ at $1\mathrm{\upmu m}$ and 1.5\,d for the lanthanide-free portion of the ejecta~\cite{Bulla:2019muo}. 
A second spherical component represents the ejecta released from the merger remnant and debris disk, extending from minimal velocities  \mbox{$v_\mathrm{min}^\mathrm{pm}=0.025$\,c} 
up to maximal velocities \mbox{$v_\mathrm{max}^\mathrm{pm}=0.08$\,c} and with an ejecta mass $M_\mathrm{ej}^\mathrm{pm}$. 
The bound-bound opacities adopted for the postmerger ejecta are intermediate  \cite{Perego:2017wtu} to those in the lanthanide-rich and lanthanide-free components of the dynamical ejecta ($\kappa_\mathrm{bb}=0.1$\,cm$^2$ g$^{-1}$ at $1\mathrm{\upmu m}$ and 1.5\,d). SEDs and corresponding light curves are then controlled by four parameters: 
$M_\mathrm{ej}^\mathrm{dyn}$, $M_\mathrm{ej}^\mathrm{pm}$, 
$\Phi$, and the observer viewing angle $\Theta_\mathrm{obs}$.

\textit{Model II:} This model adopts a kilonova without an additional wind ejecta component~\cite{Bulla:2019muo}, 
which makes standardization and extraction of the Hubble constant easier 
due to the smaller number of free parameters. 
Tighter constraints on the distance and inclination angle are extracted compared to our 
standard choice (Model I); cf.~Fig.~\ref{fig:D_suppl}. 

\textit{Model III:}
This model adopts the radiative transfer model 
of Ref.~\cite{Kasen:2017sxr} and employs a multi-dimensional 
Monte Carlo code to solve the multi-wavelength radiation transport 
equation for an expanding medium. 
We use one spherically symmetric ejecta component characterized by 
the mass of the ejecta $M_{\rm ej}$, 
the mass fraction of lanthanides $X_{\rm lan}$, 
and the ejecta velocity $v_{\rm ej}$. 
While using only one ejecta component reduces the consistency between 
the observational data and the model prediction, 
it provides easier standardization and therefore puts a tighter constraint 
on the measured distance, 
but no information about inclination can be extracted due to the assumption of spherical symmetry. 

\textit{Surrogate Construction:}
We use the approach outlined in Refs.~\cite{Doctor:2017csx,Purrer:2014fza}, 
where a Gaussian-Process-Regression framework is employed; 
cf. Refs.~\cite{Coughlin:2018miv,Coughlin:2019zqi} for a detailed discussion.\\

\begin{figure}[h!]
    \centering
    \includegraphics[width=0.9\textwidth]{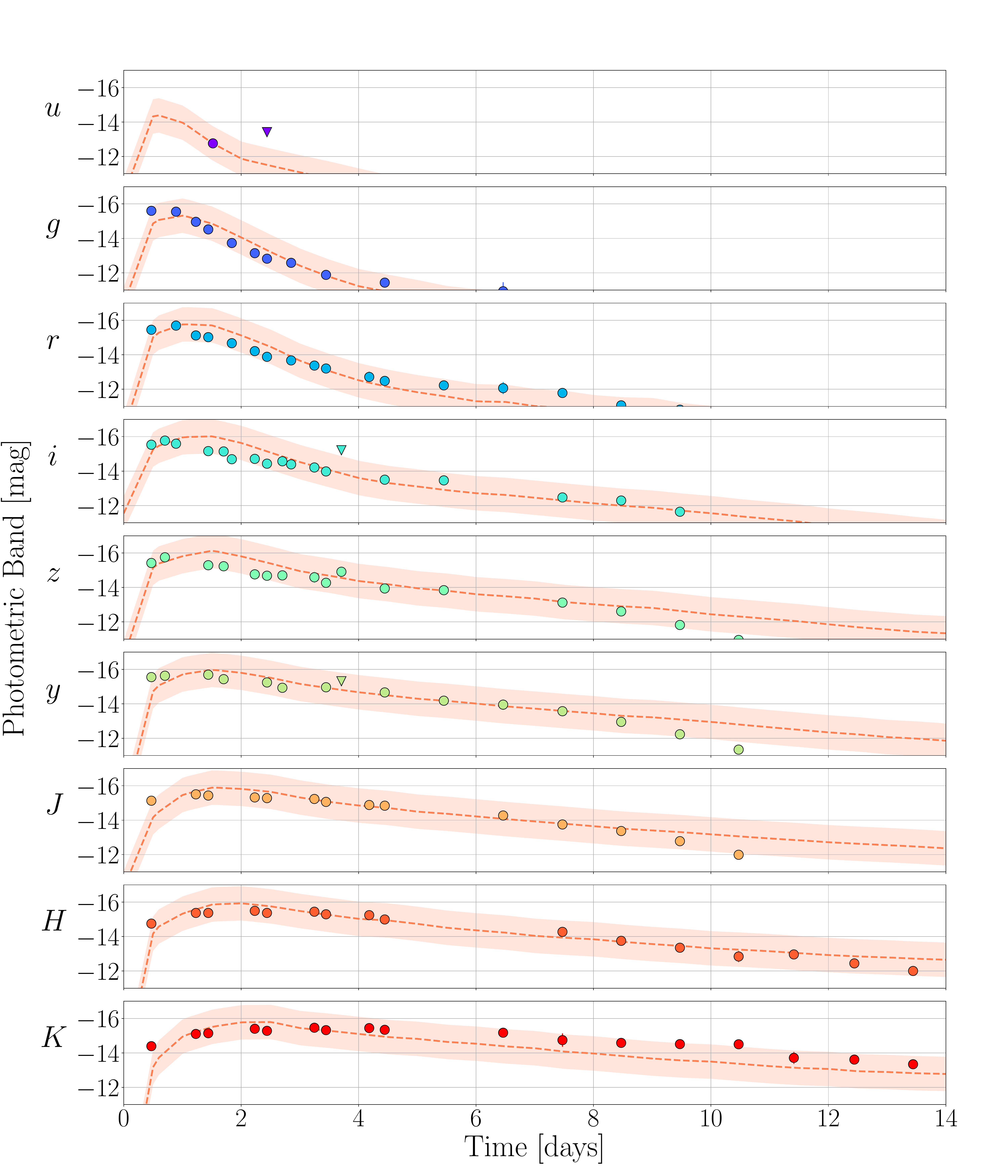}
    \caption{\textbf{Comparison of observed light curves of AT2017gfo with Model I.} 
    Predictions of Model I (shaded bands) are compared to observational data (points) in different photometric bands collected in~\cite{Coughlin:2018miv} 
    using the original data of~\cite{Andreoni:2017ppd,Arcavi:2017xiz,Chornock:2017sdf,Cowperthwaite:2017dyu,
    Drout:2017ijr,Evans:2017mmy,Kasliwal:2017ngb,Tanvir:2017pws,Pian:2017gtc,Troja:2017nqp,
    Smartt:2017fuw,Utsumi:2017cti,Valenti:2017ngx}.}
    \label{fig:EM_model}
\end{figure}

\begin{figure}[h!]
    \centering
    \includegraphics[width=0.8\textwidth]{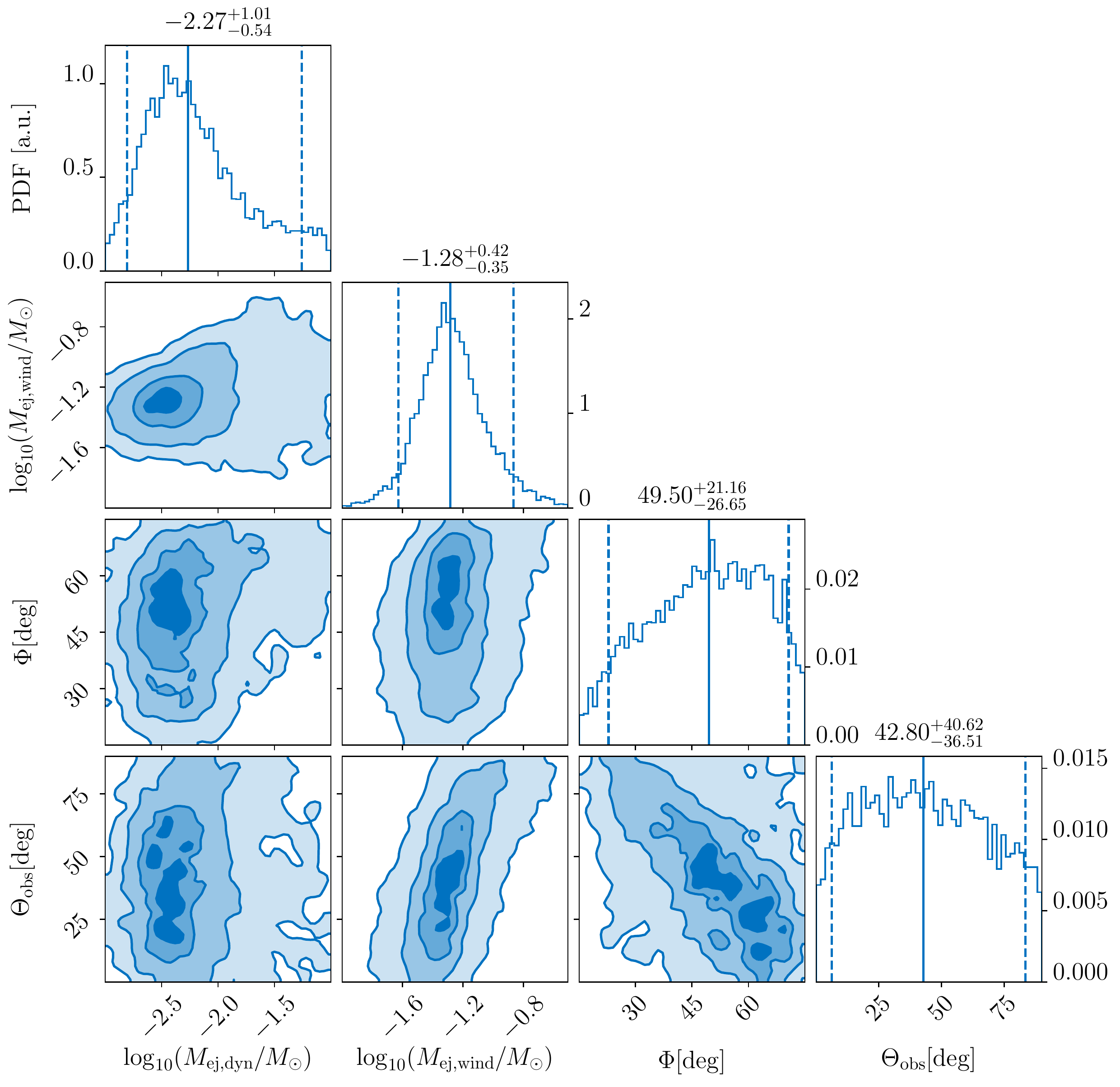}  
    \caption{\textbf{Estimated ejecta properties for Model I.} Corner plot for the mass of the dynamical ejecta $M_{\rm ej, dyn}$, the mass of the disk wind ejecta $M_{\rm ej,wind}$, 
    the opening angle between lanthanide-rich and lanthanide-poor dynamical ejecta components $\Phi$, and the viewing angle $\Theta_{\rm{obs}}$ at 10\%, 32\%, 68\% and 95\% confidence.
    For the 1D posterior probability distributions, we mark the median (solid lines) and the 90\% confidence interval (dashed lines) and report these above each panel.}
    \label{fig:ejecta_par_corner}
\end{figure}

We show the performance of our standard model (Model I) in Fig.~\ref{fig:EM_model}
and find that it is consistent with the observed data. 
The extracted properties of the ejecta are shown in Fig.~\ref{fig:ejecta_par_corner}. 
The disk wind ejecta are about 10 times larger than the dynamical ejecta.
The angle $\Phi$ peaks around $50^\circ$, while the observation angle $\Theta_{\rm{obs}}$ peaks around $40^\circ$ (cf.~Fig.~\ref{fig:Bulla_geometry}).

To connect the individual ejecta components to the different ejecta mechanisms, 
we assume that the total ejecta mass is a sum of multiple components.
The first component is related to the dynamical ejecta $ M^{\rm dyn}_{\rm ej}$.
The second component is caused by disk wind ejecta and proportional to the 
disk mass surrounding the final remnant $M_{\rm ej}^{pm} = \zeta \ M_{\rm disk}$.
For a conservative estimate, we also add a third component $\alpha$ that we 
keep as a free parameter during the sampling procedure. 

For the dynamical ejecta, we use the description in Ref.~\cite{Coughlin:2018fis}, while we assume that the 
disk wind ejecta is proportional to the disk mass. 
Based on recent works on predicting 
the disk mass for systems with high mass ratios~\cite{Kiuchi:2019lls}, we include an explicit mass-ratio dependence as described below. 

\begin{figure}[h!]
    \centering
    \includegraphics[width=0.9\textwidth]{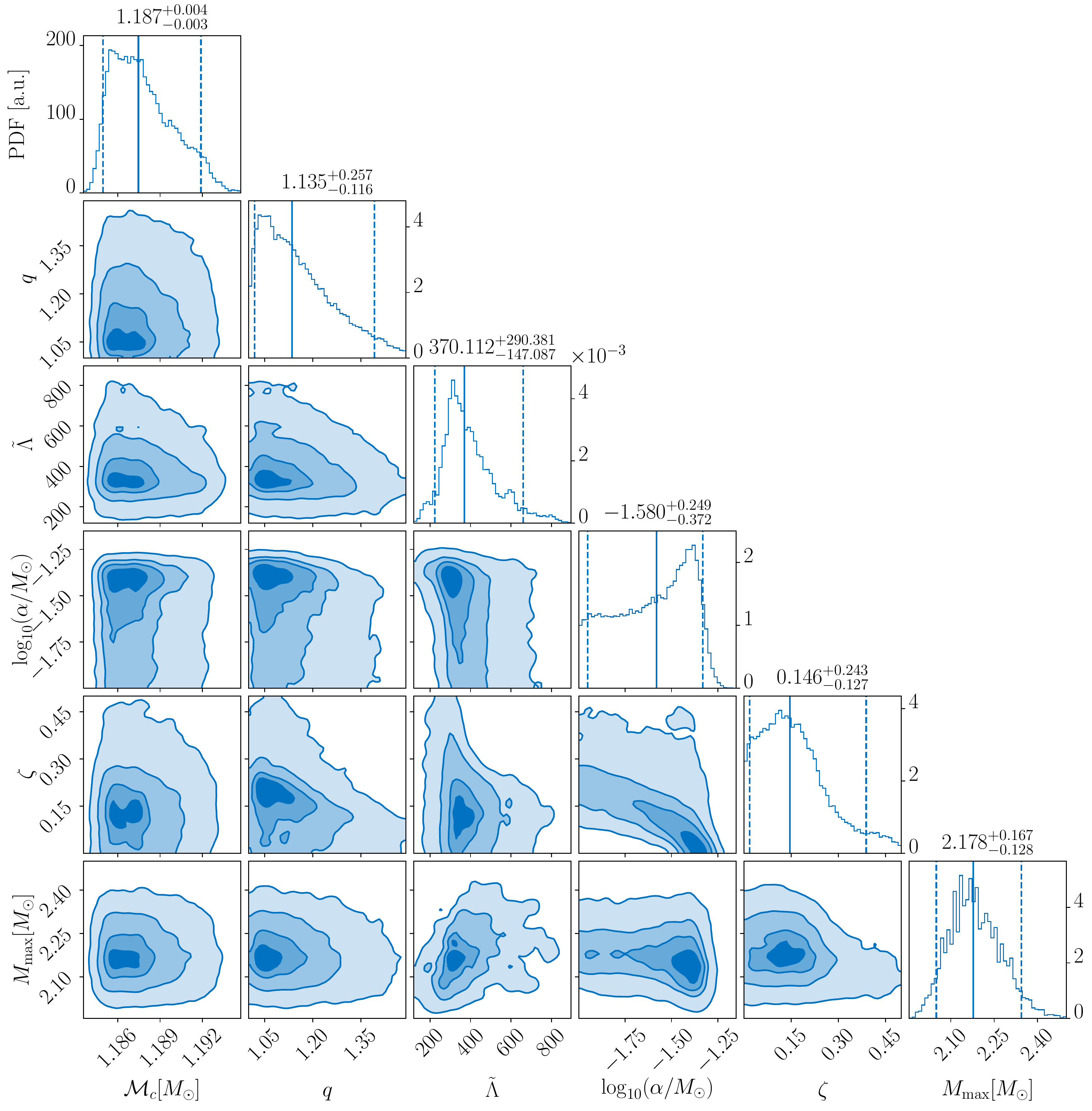}  
    \caption{\textbf{Estimated BNS properties for Model I.} Similar to Fig.~\ref{fig:ejecta_par_corner} but for the chirp mass $\mathcal{M}_c$, mass ratio $q$, tidal deformability $\tilde{\Lambda}$, free ejecta parameter $\alpha$, 
    disk conversion factor $\zeta$, and maximum NS mass. 
    Model II and Model III provide very similar binary properties.}
    \label{fig:ejecta_corner}
\end{figure}

The extracted binary properties are shown in Fig.~\ref{fig:ejecta_corner}, 
in which we report the chirp mass, the mass ratio, 
the deformability $\tilde{\Lambda}$, 
the fraction of the dynamical ejecta $\alpha$, the disk conversion factor $\zeta$, 
and the maximum TOV mass. 

\subsubsection*{Disk mass prediction}

\begin{figure}[t]
    \centering
    \includegraphics[width=0.48\textwidth]{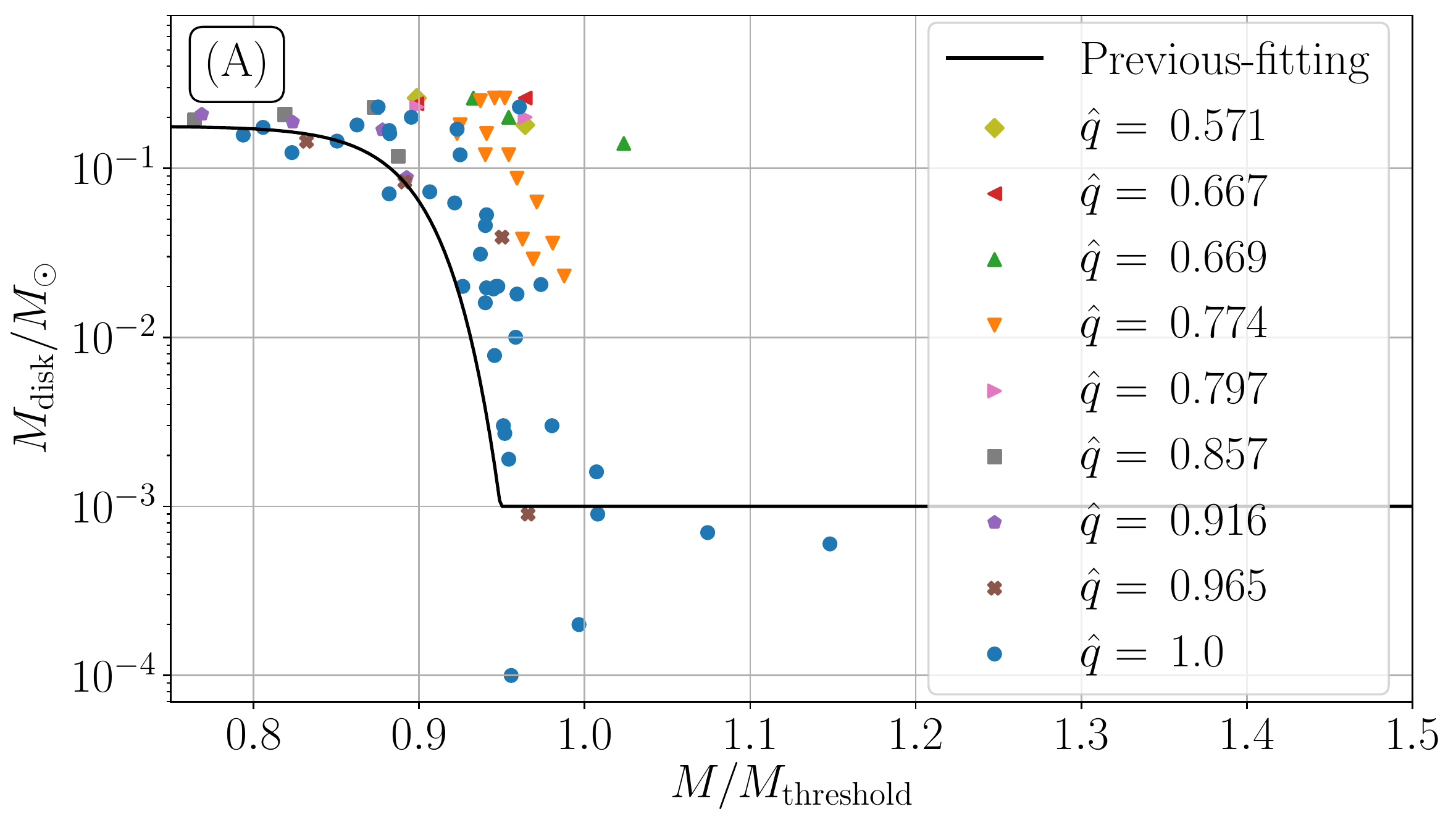}
    \includegraphics[width=0.48\textwidth]{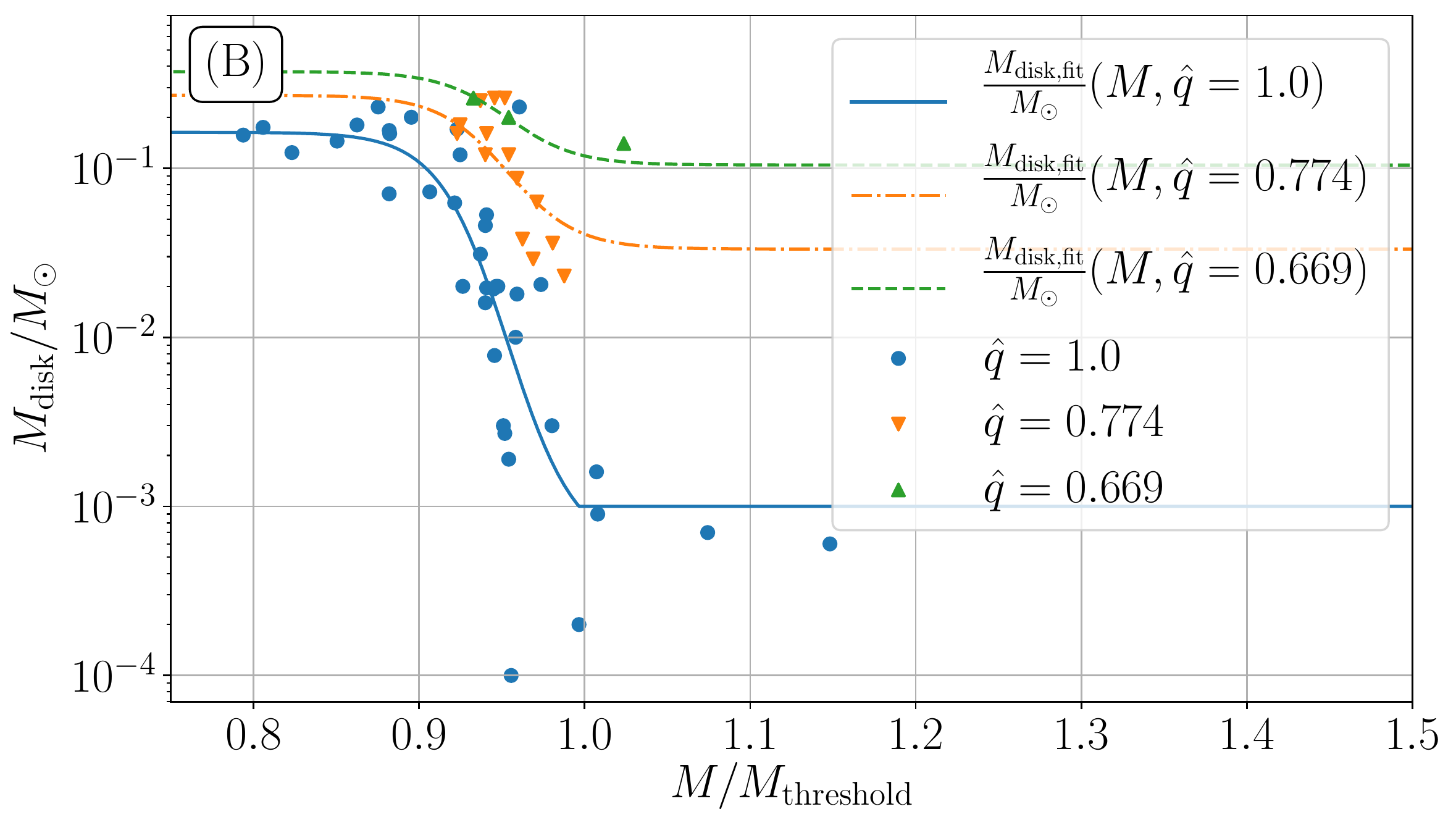}    
    \caption{{\bf Disk mass predictions for various total masses and mass ratios.} (A) Data employed for the construction of the model fitting in Eq.~\eqref{eq:Mdisk_fit}, compared to the model of~\cite{Coughlin:2018fis}.
    (B) Three examples, where data from numerical-relativity simulations (symbols) are compared to the fit for different mass ratios (lines).
    }
    \label{fig:disk_mass}
\end{figure}

We utilise results from 73 numerical relativity 
simulations performed by different groups~\cite{Kiuchi:2019lls,Radice:2018pdn,Dietrich:2016hky,Hotokezaka:2011dh}. 
The full dataset is shown in Fig.~\ref{fig:disk_mass}A which shows the 
disk mass versus the ratio of the total mass of the system and the threshold mass.
The threshold mass $M_{\rm threshold}$ is the limiting total mass of the BNS system beyond which a prompt collapse to a black hole occurs.
For the estimate of the threshold mass, 
we use the predictions of Ref.~\cite{Agathos:2019sah}.
We compare the data with the estimate of Ref.~\cite{Coughlin:2018fis} confirming that an increasing mass ratio leads to an increased disk mass~\cite{Kiuchi:2019lls}. 
We use a similar functional behavior to Ref.~\cite{Coughlin:2018fis}, but
we incorporate mass-ratio dependent fitting parameters such that
\begin{equation}
    \log_{10}\left(\frac{M_{\rm disk}}{M_{\odot}}\right) = \textrm{max}\left(-3, a\left(1+b\tanh\left(\frac{c - (m_1+m_2) M_{\rm threshold}^{-1}}{d}\right)\right)\right), \label{eq:Mdisk_fit}
\end{equation}
with $a$ and $b$ given by
\begin{equation}
\begin{aligned}
    a &= a_o + \delta a \cdot \xi\,,\\
    b &= b_o + \delta b \cdot \xi\,,
\end{aligned}
\end{equation}
where $a_o$, $b_o$, $\delta a$, $\delta b$, $c$, and $d$ are free parameters. The parameter $\xi$ is given by
\begin{equation}
    \xi = \frac{1}{2}\tanh\left(\beta \left(\hat{q}-\hat{q}_{\rm trans}\right)\right)\,,
\end{equation}
where $\hat{q} \equiv m_2/m_1 \leq 1$ is the inverse mass ratio and $\beta$ and $\hat{q}_{\rm trans}$ are free parameters.
Fig.~\ref{fig:disk_mass}B 
shows how the model fitting changes as the mass ratio changes. 

The best-fitting model parameters are given by minimizing $r = \langle(\log_{10}(M_{\rm disk}) - \log_{10}(M^{\rm fit}_{\rm disk}))^2\rangle$; 
we find $a_o=-1.581$, $\delta a=-2.439$, $b_o=-0.538$, $\delta b=-0.406$, $c =0.953$, $d=0.0417$, $\beta=3.910$, $\hat{q}_{\rm trans}=0.900$.

\subsection*{Prior combination for distance measurement}
Due to the strong correlation between the luminosity distance $D$ and inclination 
$\iota_0$ across different analyses, 
we combine the information on the $D$-$\iota_0$ plane and then marginalize over the inclination.
We take the GRB170817A-VLBI measurement 
$p_{\textrm{GRB}}(D,\iota_0)$ as the prior for the other two analyses. 
Therefore, the combined posterior probability distribution $p_{\textrm{com}}(D,\iota_0)$ is given by
\begin{equation}
	p_{\textrm{com}}(D,\iota_0) = \mathcal{L}_{\textrm{GW}}(D,\iota_0) \times \mathcal{L}_{\textrm{EM}}(D,\iota_0) \times p_{\textrm{GRB}}(D,\iota_0),
\end{equation}
where $\mathcal{L}_{\textrm{GW}}$ and $\mathcal{L}_{\textrm{EM}}$ are the 
likelihoods for the parameters $(D,\iota_0)$ for the GW170817 and AT2017gfo analyses, respectively.

Because we are combing the information in the post-processing stage, 
we do not have access to the likelihood but only the posterior probability distributions of GW170817, $p_{\textrm{GW}}$, 
and AT2017gfo, $p_{\textrm{EM}}$. Therefore, we evaluate the combined posterior probability distribution by
\begin{equation}
	p_{\textrm{com}}(D,\iota_0) = \frac{p_{\textrm{GW}}(D,\iota_0)}{\pi_{\textrm{GW}}} \times \frac{p_{\textrm{EM}}(D,\iota_0)}{\pi_{\textrm{EM}}} \times p_{\textrm{GRB}}(D,\iota_0),
\end{equation}
where $\pi_{\textrm{GW}}$ and $\pi_{\textrm{EM}}$ are the priors for the parameters $(D,\iota_0)$ used for analysing GW170817 and AT2017gfo, respectively.

The combined posterior probability distribution on the distance is then given by
\begin{equation}
	p_{\textrm{com}}(D) = \int d\iota_0 \, p_{\textrm{com}}(D,\iota_0)
\end{equation}
which we use below in the Hubble constant measurement. 

\subsection*{Estimation of the Hubble constant $H_0$}
The Hubble constant $H_0$ relates the center-of-mass recession velocity of a galaxy relative to the cosmic microwave background (CMB)~\cite{Hinshaw:2008kr} $v_r$ with the comoving distance $D_c$ and the peculiar velocity $v_p$ by
\begin{equation}
	v_r = H_0D_c + v_p\,.
\end{equation}
The distance between Earth and NGC~4993, the host galaxy of GW170817, is small, $40~\textrm{Mpc}$~\cite{Coulter:2017wya},
so we can approximate the comoving distance with the luminosity distance $D$.
Combining the distance measurement with the redshift $z$ of the host galaxy, $z~=~0.009783~\pm~0.000023$, constrains the Hubble constant~\cite{Abbott:2017xzu}.

GW170817's host galaxy NGC~4993 belongs to the galaxy cluster ESO~508, which has a radial velocity of $v_r$ of $3327\pm72$km s$^{-1}$~\cite{Crook:2006sw} and the peculiar velocity $v_p$ of NGC~4993 is $310\pm69$~km s$^{-1}$~\cite{Springob:2014qja} . 
To reduce possible systematics introduced by imperfect modelling of the bulk flow motion~\cite{Springob:2014qja}, 
we take the uncertainty on $v_p$ to be $150$ km s$^{-1}$~\cite{Carrick_2015}. 

We model the likelihoods of $v_r$, $\mathcal{L}(v_r)$, and $v_p$, $\mathcal{L}(v_p)$, to be Gaussians given by
\begin{equation}
	\mathcal{L}(v_r) \propto \exp\left(-\frac{1}{2}\left(\frac{v_r-\langle v_r \rangle}{\sigma_{v_r}}\right)^2\right), \quad 
	\mathcal{L}(v_p) \propto \exp\left(-\frac{1}{2}\left(\frac{v_p-\langle v_p \rangle}{\sigma_{v_p}}\right)^2\right),
\end{equation}
where $\langle v_r \rangle=3327$~km s$^{-1}$, $\sigma_{v_r}=72$~km s$^{-1}$, $\langle v_p \rangle=310$~km s$^{-1}$ and $\sigma_{v_p}=150$~km s$^{-1}$.

As a result, the multi-dimensional posterior probability distribution $p(H_0, D, v_p)$ is given by
\begin{equation}
\begin{aligned}
	p(H_0, D, v_p) &= \mathcal{L}(H_0, D, v_p)\pi(H_0, D,v_p) \times \frac{1}{\mathcal{N}_s(H_0)}\\
	              &\propto \exp\left(-\frac{1}{2}\left(\frac{v_p-\langle v_p \rangle}{\sigma_{v_p}}\right)^2\right) \times \exp\left(-\frac{1}{2}\left(\frac{H_0D + v_p-\langle v_r \rangle}{\sigma_{v_r}}\right)^2\right)\\
		      &\times p(D) \times \pi(H_0) \times \pi(v_p)\times \frac{1}{\mathcal{N}_s(H_0)},
\end{aligned}
\end{equation}
where $p(D)$, $\pi(H_0)$ and $\pi(v_p)$ are the posterior probability distribution of the distance, the prior on the Hubble constant, and the prior on the peculiar velocity, respectively. 
$\mathcal{N}_s(H_0)$ is the selection effect term~\cite{Abbott:2017xzu}. 
We take $\pi(H_0)$ to be uniform in $[20,160]~\textrm{km s}^{-1}\textrm{Mpc}^{-1}$, $\pi(v_p)$ to be uniform in $[-c,c]$ and $\mathcal{N}_s(H_0) \propto H^3_0$.
This choice of selection effect term is rooted in a volumetric prior on the redshift~\cite{Abbott:2017xzu}.

\begin{figure}[h!]
    \centering
    \includegraphics[width=0.5\textwidth]{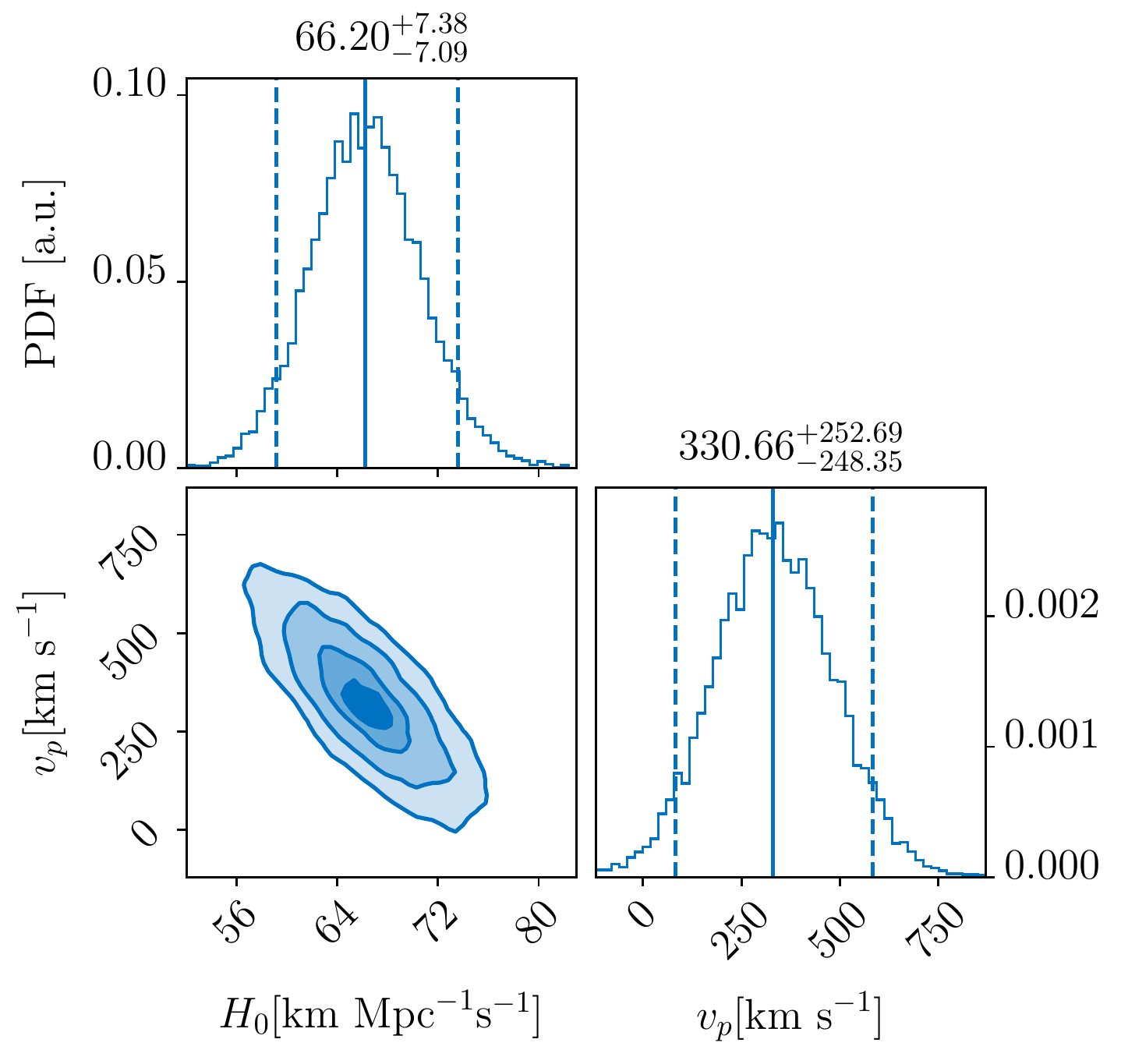}
    \caption{\textbf{Constraints on the Hubble constant.} 
    Corner plot of the inferred $H_0$-$v_p$ posterior probability distribution using the inferred distance from our analysis, cf. Fig.~\ref{fig:D_i0}. For the 1D posterior probability distributions, we mark the median (solid lines) and the 90\% confidence interval (dashed lines) and report these above each panel.}
    \label{fig:H_0_v_p}
\end{figure}

For the posterior probability distribution of the distance, we take the posterior probability distribution based on the combined analysis as described above, including the use of standardizable kilonovae light curves to measure their distances~\cite{Coughlin:2019vtv,Kashyap:2019ypm}. 
Because we have a set of posterior probability distribution samples $\{d_i\}$ that follow the posterior probability distribution $p_\textrm{com}(D)$, 
we obtain the marginalized posterior probability distribution $p(H_0,v_p)$ by
\begin{equation}
\begin{aligned}
	p(H_0,v_p) &= \int dD \, p(H_0,D, v_p)\\
	&= \int dD\, p_\textrm{com}(D) \frac{p(H_0, D, v_p)}{p_\textrm{com}(D)}\\ 
	&= \left\langle \frac{p(H_0, D, v_p)}{p_\textrm{com}(D)} \right\rangle_{\{d_i\}},
\end{aligned}
\end{equation}
in which we approximate $\int dD \,p_\textrm{com}(D)$ by an average over posterior probability distribution samples denoted $\langle \cdots \rangle_{\{d_i\}}$. 
We sample over $p(H_0,v_p)$ with \textsc{emcee}~\cite{ForemanMackey:2012ig} and obtain the corner plot shown in Fig.~\ref{fig:H_0_v_p}.

\clearpage 

\section*{Supplementary Text}

\subsection*{Gravitational-Wave Analysis}

\begin{figure}[h!]
    \centering
    \includegraphics[width=0.9\textwidth]{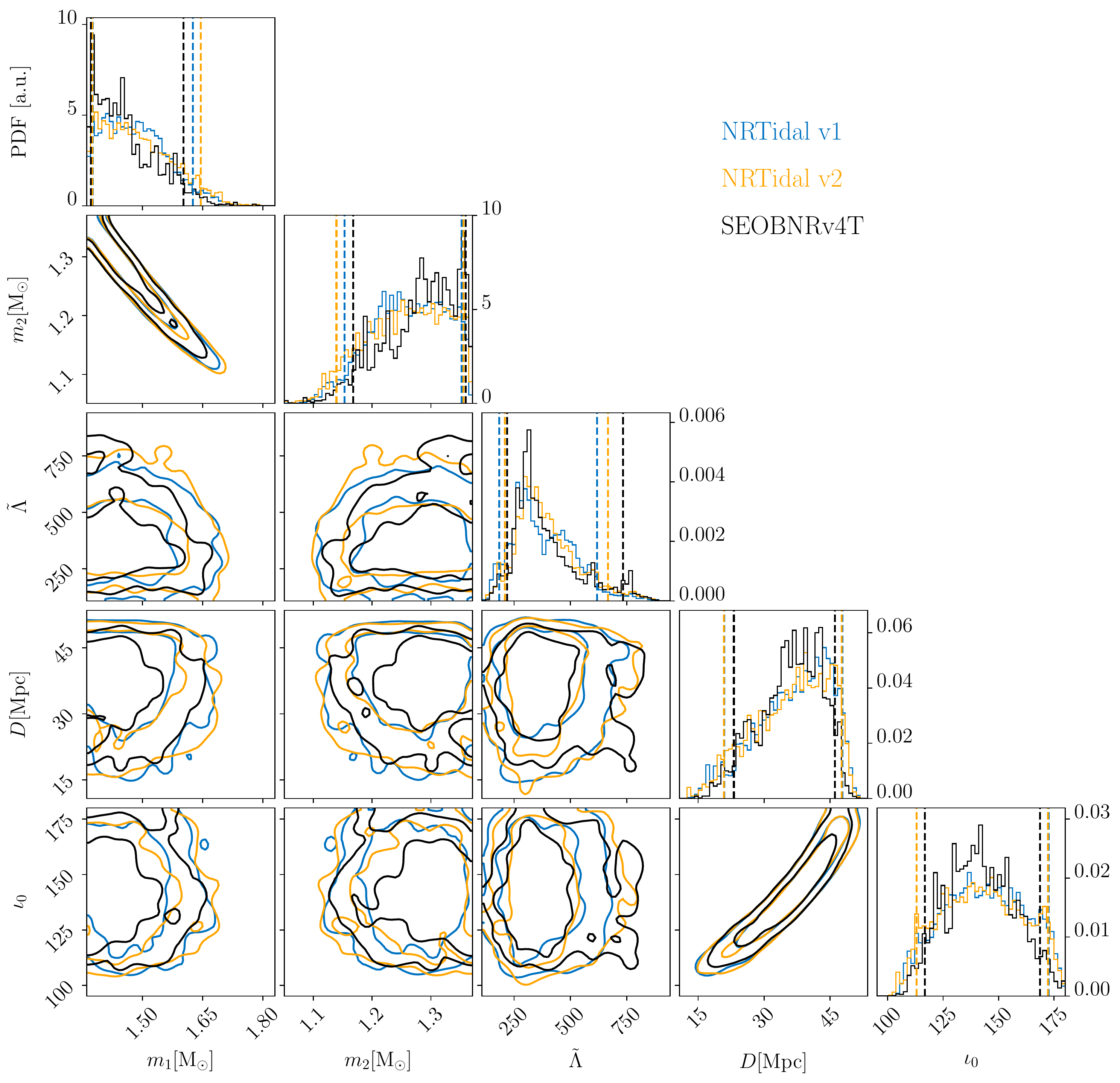}
	\caption{\textbf{Marginalized 1D and 2D posterior probability distributions of GW170817's parameters.}
	Visualization of the 1D and 2D marginalized posterior probability distributions (corner plot) of the parameters of GW170817 at 68\% and 95\% confidence, obtained with \texttt{IMRPhenomPv2\_NRTidal} (blue), \texttt{IMRPhenomPv2\_NRTidalv2} (orange) and \texttt{SEOBNRv4T} (black) for the primary source mass $m_1$, secondary source mass $m_2$, mass-weighted tidal deformability $\tilde{\Lambda}$, luminosity distance $D$, and inclination $\iota_0$. 
	For the 1D posterior probability distributions, we show the probability distribution function  (PDF) in arbitrary units (a.u.) and mark the 90\% confidence confidence interval by dashed lines.
	The main difference between the posterior probability distributions inferred with the three waveform models is the distribution of $\tilde{\Lambda}$, which is expected due to the different tidal description of the three models.} 
    \label{fig:GW170817_corner}
\end{figure}

As discussed in the main text, we use the \texttt{IMRPhenomPv2\_NRTidalv2} (\texttt{NRTidalv2}) waveform model. 
The approximant uses the description of tidal effects introduced in Ref.~\cite{Dietrich:2019kaq} to augment the precessing binary black-hole waveform model~\cite{Hannam:2013oca}. 
\texttt{NRTidalv2} has a different tidal and spin description to the 
\texttt{IMRPhenomPv2\_NRTidal} model~\cite{Dietrich:2017aum,Dietrich:2018uni}, which was the waveform model employed by the 
LIGO Scientific and Virgo Collaborations to interpret GW170817~\cite{Abbott:2018wiz,Abbott:2018exr,LIGOScientific:2019eut,
LIGOScientific:2018mvr,Abbott:2018lct} 
and GW190425~\cite{Abbott:2020uma}. 
We present the parameter-estimation results for GW170817 in Fig.~\ref{fig:GW170817_corner}. 
For comparison, we also show the posterior probability distributions obtained with the \texttt{IMRPhenomPv2\_NRTidal} 
waveform model to allow for an assessment of \texttt{NRTidalv2}.
We find no noticeable difference in the measured component masses, distance, and inclination; cf. Tab.~\ref{tab:GW170817_waveform_models_suppl}. 
The agreement is likely caused by the same underlying point-particle base line of both models. 
There is a small difference in the estimated tidal deformability, where \texttt{NRTidalv2}
predicts a slightly larger tidal deformability which consequently results in a slightly larger radius estimate. 
This behavior is expected because the \texttt{NRTidalv2} approximant incorporates slightly smaller tidal contributions 
for the same physical parameters 
than the original \texttt{IMRPhenomPv2\_NRTidal} model, which consequently leads to a larger estimated tidal deformability. 
In addition, we show posterior probability distributions obtained with the \texttt{SEOBNRv4T}
waveform model~\cite{Hinderer:2016eia}, where we employ its surrogate model of Ref.~\cite{Lackey:2018zvw} for the parameter estimation runs. 
\texttt{SEOBNRv4T} has a point-particle and tidal description that differs from \texttt{IMRPhenomPv2\_NRTidal} 
and \texttt{NRTidalv2} and, therefore, provides an independent check for possible systematic uncertainties.
We find no noticeable difference between parameters, 
suggesting that no systematic errors are introduced by the choice of waveform model in our analysis; cf.~Fig.~\ref{fig:GW170817_corner} and Tab.~\ref{tab:GW170817_waveform_models_suppl}.

\begin{table}[h!]
\centering
\caption{\textbf{Summary of the parameters of GW170817 inferred with different waveform models}. We give the median of the parameters of GW170817, together with their corresponding $90\%$ credible intervals for analyses using different waveform models.}
\begin{tabular}{l|l|l|l}
        Parameter & \texttt{NRTidal} & \texttt{NRTidalv2} & \texttt{SEOBNRv4T}\\
        \hline
	Primary mass $m_1 \ [M_{\odot}]$ & $1.48^{+0.15}_{-0.10}$ & $1.48^{+0.17}_{-0.10}$ & $1.45^{+0.16}_{-0.07}$\\
	Secondary mass $m_2 \ [M_{\odot}]$ & $1.26^{+0.09}_{-0.11}$ & $1.26^{+0.09}_{-0.12}$ & $1.29^{+0.07}_{-0.12}$\\
	Mass-weighted tidal deformability $\tilde{\Lambda}$ & $357.86^{+259.49}_{-173.26}$ & $370.54^{+296.33}_{-160.57}$ & $349.25^{+383.87}_{-129.77}$\\
	Luminosity distance $D \ [\rm{Mpc}]$ & $37.85^{+9.95}_{-16.95}$ & $37.33^{+10.29}_{-16.46}$ & $36.81^{+9.30}_{-13.69}$\\
	Inclination $\iota_0 \ [\rm{deg}]$& $143.41^{+29.07}_{-30.31}$ & $142.19^{+30.08}_{-29.15}$ & $141.21^{+27.48}_{-24.48}$\\
\end{tabular}
\label{tab:GW170817_waveform_models_suppl}
\end{table}

\begin{figure}[h!]
    \centering
    \includegraphics[width=0.9\textwidth]{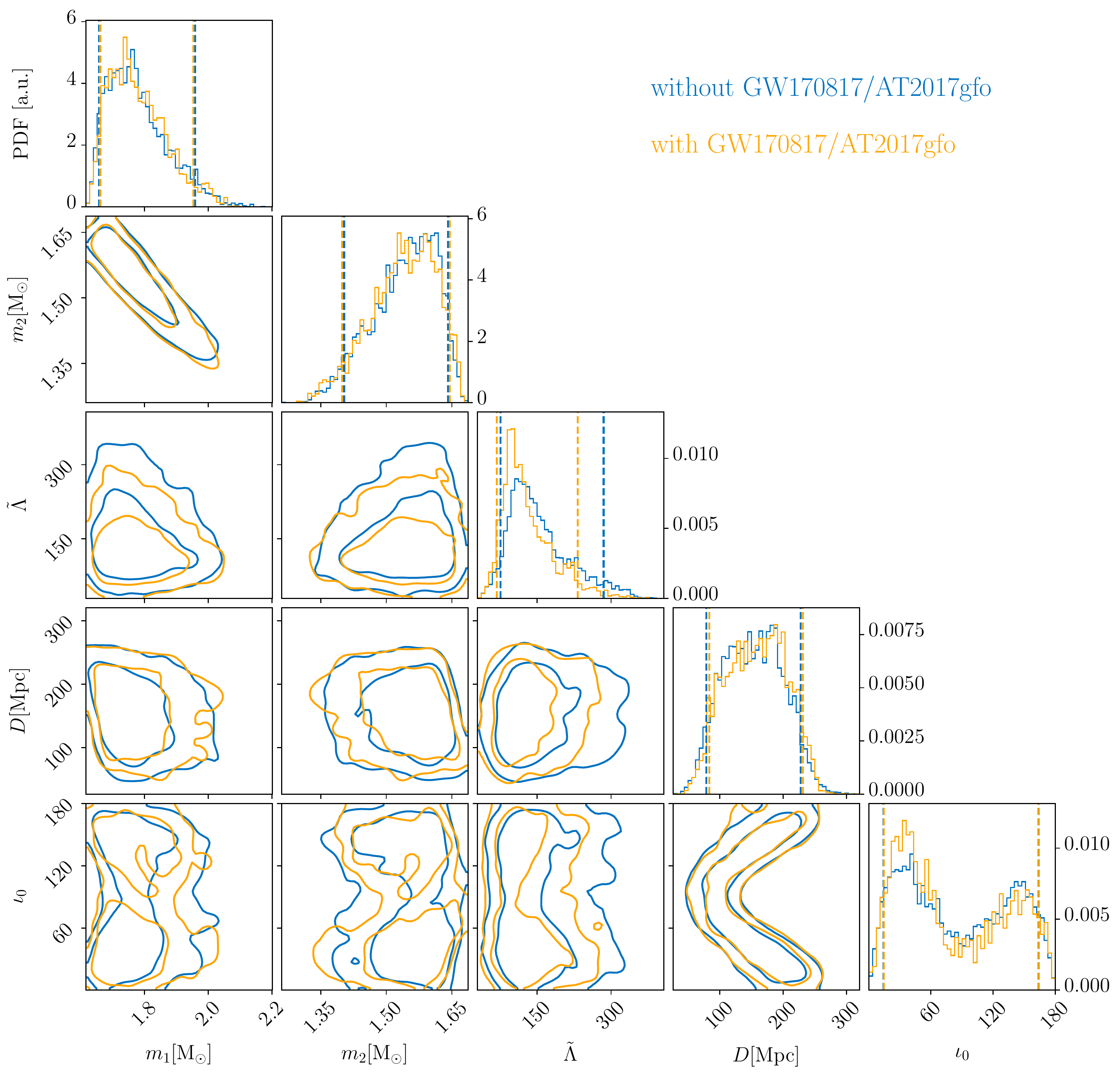}
	\caption{\textbf{Marginalized 1D and 2D posterior probability distributions of GW190425's parameters.}
	Same as Fig.~\ref{fig:GW170817_corner} but for the posterior probability distribution of GW190425's parameters obtained without (blue) and with (orange) the inclusion of GW170817/AT2017gfo. 
	With the inclusion of GW170817/AT2017gfo, the tidal measurement is more tightly constrained.}
    \label{fig:GW190425_corner}
\end{figure}

For the analysis of GW190425 we also use the \texttt{NRTidalv2} model. 
In Fig.~\ref{fig:GW190425_corner}, we show extracted source parameters for GW190425 
when we include or do not include the GW170817 and AT2017gfo information (chirp mass, mass ratio, and EOS constraints); cf.~Tab.~\ref{tab:GW190425_with_without}.
The extracted source parameters differ in the estimated tidal deformability $\tilde{\Lambda}$, and the inclusion of GW170817 and AT2017gfo in our analysis leads to a smaller value. 
The incorporation of additional information from GW170817 and AT2017gfo changes the prior of the GW190425 analysis such that NSs with large radii (large tidal deformabilities) are 
already disfavored.

\begin{table}[h!]
\centering
\caption{\textbf{Summary of the parameters of GW190425 with and without inclusion of GW170817/AT2017gfo}. 
We give the median of the parameters of GW190425, together with their corresponding $90\%$ credible intervals for analyses with and without input from GW170817/AT2017gfo.}
\begin{tabular}{p{6.6cm}|p{3.85cm}|p{3.85cm}}
        Parameter & without GW170817/AT2017gfo & with GW170817/AT2017gfo\\
        \hline
	Primary mass $m_1 \ [M_{\odot}]$ & $1.76^{+0.20}_{-0.11}$ & $1.77^{+0.19}_{-0.10}$\\
	Secondary mass $m_2 \ [M_{\odot}]$ & $1.55^{+0.09}_{-0.15}$ & $1.54^{+0.10}_{-0.15}$\\
	Mass-weighted tidal deformability $\tilde{\Lambda}$ & $140.80^{+144.22}_{-64.73}$ & $117.90^{+114.60}_{-49.24}$\\
	Luminosity distance $D \ [\rm{Mpc}]$ & $152.87^{+74.52}_{-73.99}$ & $159.08^{+71.91}_{-75.73}$\\
	Inclination $\iota_0 \ [\rm{deg}]$& $79.22^{+84.81}_{-64.67}$ & $64.54^{+99.66}_{-49.88}$\\
\end{tabular}
\label{tab:GW190425_with_without}
\end{table}

\subsection*{Modelling of AT2017gfo}

\begin{figure}[h!]
    \centering
    \includegraphics[width=0.65\textwidth]{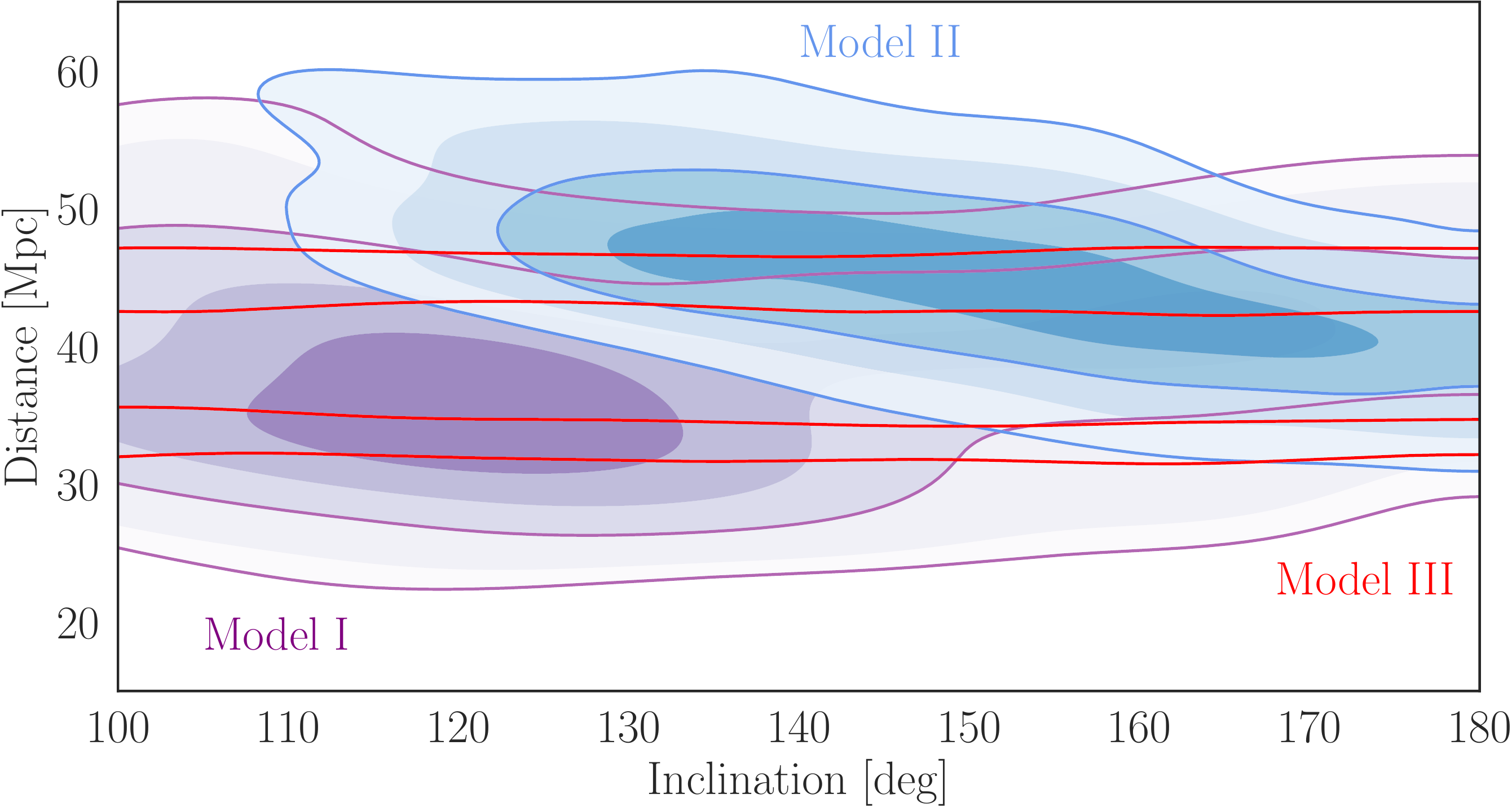}
    \caption{\textbf{Distance-inclination measurements for kilonova models.}
    Shown are results for Model I (purple), Model II (blue), and Model III (red) at 68\% and 95\% confidence levels.
    Model I is the least constraining.}
    \label{fig:D_suppl}
\end{figure}

There is good agreement between the three kilonova models, which differ mostly in the predicted inclination and distance.  
We show the distance-inclination measurements in Fig.~\ref{fig:D_suppl}. 
Model I is the least constraining due to the additional complexity of the wind ejecta component.
Model III is spherically symmetric and therefore only enables a distance measurement. 
All three models agree within their statistical uncertainties, which suggests that our analysis is dominated by statistical effects and not systematics. 

For the analysis of the non-observed EM counterpart of GW190425, 
we use the same kilonova analysis as discussed above, but restricted to Model I.
We use information from the Asteroid Terrestrial-impact Last Alert System (ATLAS)~\cite{2019GCN.24197....1M} that covered 37\%, the Gravitational-wave Optical Transient Observer (GOTO)~\cite{Gompertz:2020cur} that covered 30\%, the Master Global Robotic Telescopes Net (MASTER)~\cite{2019GCN.24167....1L} that covered 37\%, and the Zwicky Transient Facility (ZTF)~\cite{Coughlin:2019xfb} that covered 25\% of the sky area derived from the GW data to obtain apparent magnitude limits on potential counterparts from optical surveys. 
An exact computation of the total sky coverage is not possible because not all groups released their 
covered tiles and search information. 
However, the published limits, together with the distance information from the GW event, lead to 
limits on the absolute magnitude of a potential kilonova~\cite{Coughlin:2019zqi}. 
Accounting for the distance of the transient, we rule out all ejecta parameters for which the predicted magnitude would exceed the obtained apparent magnitude limit.

\subsection*{Ordering of the analysis steps}

\begin{figure}[h!]
    \centering
    \includegraphics[width=0.6\textwidth]{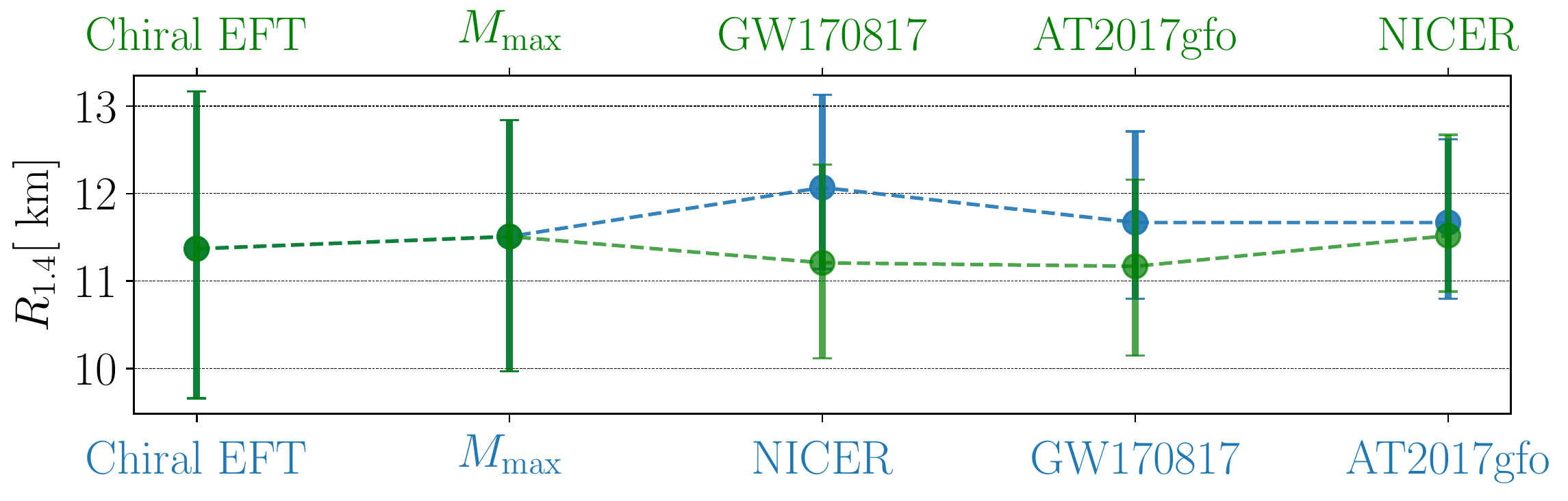}
	\caption{\textbf{Radius constraints under interchange of the individual analysis steps}. 
    The blue line is the same as in Fig.~\ref{fig:scheme}H.
	We show the highest probability interval of 90\% confidence and the median of the posterior probability distribution.}
    \label{fig:analysis_interchange}
\end{figure}

We test the effect of changing the order of analysis steps in Fig. 1 by moving the NICER results to the final stage. 
To reduce computational costs, we focus on the combination of GW170817, AT2017gfo, NICER, and the maximum-mass constraints. 
Fig.~\ref{fig:analysis_interchange} shows that the measured radius is slightly larger for our original analysis than for our analysis in which NICER results are included in the final step. 
This change is due to the kernel density estimation that is used to obtain the prior for our 
individual analysis steps. 
However, the 90\% confidence intervals remain unchanged. 
We conclude that our method is robust to the order of the procedure.

\subsection*{Propagation of systematic uncertainties}

\begin{figure}[h!]
    \centering
    \includegraphics[width=0.6\textwidth]{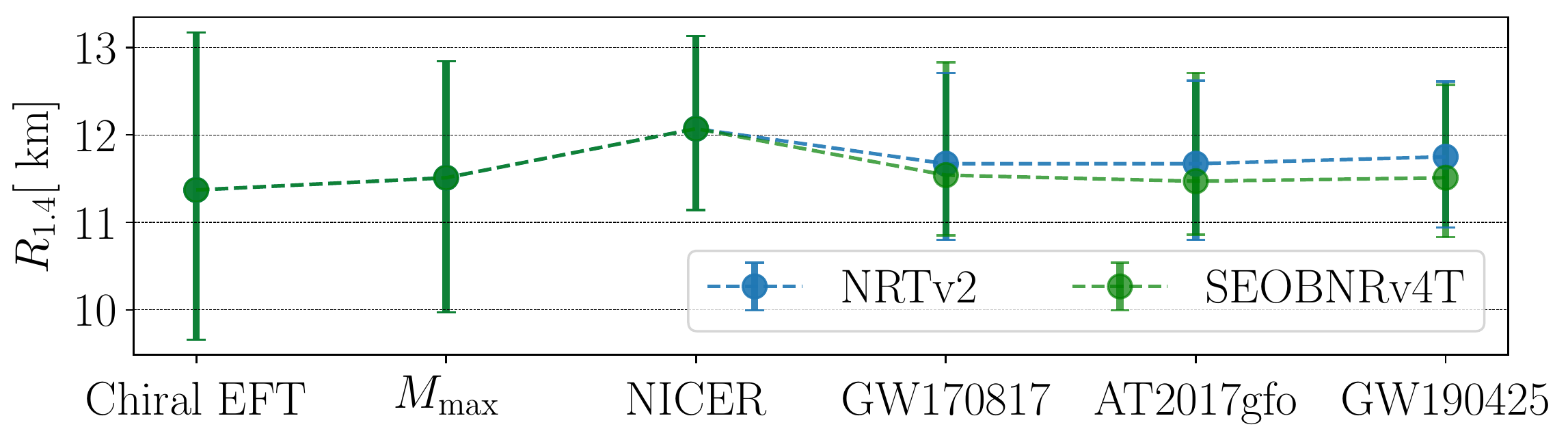}
	\caption{\textbf{Radius constraints for different gravitational-waveform models.}
	Similar to Fig.~\ref{fig:analysis_interchange} but for different gravitational-wave models used in the analyses of GW170817 and GW190425.}
    \label{fig:systematics}
\end{figure}

We also show how small differences in individual analysis steps influence the entire analysis. 
We analyse the GW events GW170817 and GW190425 with the \texttt{SEOBNRv4T} waveform model, but keep all other steps the same. 
Figure~\ref{fig:systematics} shows our final result for the radius using \texttt{IMRPhenom\_NRTidalv2} and \texttt{SEOBNRv4T}. 
\texttt{IMRPhenom\_NRTidalv2} predicts slightly larger radii than \texttt{SEOBNRv4T}
but the difference is well within uncertainties and remains almost unchanged when analysing AT2017gfo and GW190425.

Given the small sensitivity of our results to the choice of the GW model (Fig.~\ref{fig:systematics}), the order of the analysis steps (Fig.~\ref{fig:analysis_interchange}), and the consistent results employing different kilonova models (Fig.~\ref{fig:D_suppl}),
we conclude that our results are generally robust.

\end{document}